\documentclass[aps,twocolumn,showpacs,amsmath,amssymb,pra,floatfix,longbibliography,groupedaddress]{revtex4-2}

\usepackage{braket}
\usepackage{amsmath}
\usepackage{amssymb}
\usepackage{mathtools}
\usepackage{graphicx}
\usepackage{dcolumn}
\usepackage{bm}
\usepackage{multirow}
\usepackage{leftidx}
\usepackage{color}
\usepackage{float}
\usepackage{qcircuit}
\usepackage{ifthen}
\usepackage{float}
\usepackage[hidelinks]{hyperref}

\usepackage{amsfonts}
\usepackage[german,english]{babel}

\newcommand \be{\begin{equation}}
\newcommand \ee{\end{equation}}
\newcommand \bea{\begin{eqnarray}}
\newcommand \eea{\end{eqnarray}}
\newcommand \bse{\begin{subequations}}
\newcommand \ese{\end{subequations}}

\newboolean{ShowComments}
\setboolean{ShowComments}{false}  

\definecolor{mscolor}{rgb}{0,0.5,0.5}

\definecolor{jcbcolor}{rgb}{0.5,0.0,0.5}

\definecolor{rkccolor}{rgb}{0.5,0.5,0}

\begin{document}
\title{
Reducing Rydberg state dc polarizability by microwave dressing} 
\author{J.C. Bohorquez, R. Chinnarasu}

\author{J. Isaacs}
\altaffiliation{Present address: Eikon Therapeutics, Hayward, CA 94545, USA}

\author{D. Booth}
\altaffiliation{Present address: Quantum Valley Ideas Lab, Waterloo, ON N2L 6R2, Canada }

\author{M. Beck}
\altaffiliation{Present address: IBM Thomas J Watson Research Center, Yorktown Heights, NY 10598, USA}

\author{R. McDermott, and M. Saffman}
\affiliation{Department of Physics, University of Wisconsin-Madison,
Madison, Wisconsin 53706, USA}

\date{\today}

\begin{abstract}
We demonstrate reduction of the  dc polarizability of Cesium atom Rydberg states in a 77 K  environment utilizing microwave field dressing. In particular we reduce the polarizability of $52P_{3/2}$ states which have  resonances at  5.35 GHz to $51D_{5/2}$, suitable for interfacing Rydberg atoms to superconducting resonators in a cryogenic environment. We measure the polarizability of the Rydberg states using Magneto-Optical-Trap (MOT) loss spectroscopy. Using an off-resonant radio-frequency (RF) dressing field coupling $52P_{3/2}$ and $51D_{5/2}$  we  demonstrate a reduction in dc polarizability of the $ 52P_{3/2}$ states over 80$\%$.  Experimental findings are in good agreement with a numerical model of the atom-dressing field system developed using the Shirley-Floquet formalism. We also demonstrate that the dc polarizability reduction is highly anisotropic, with near total nulling possible when the dc and dressing fields are aligned, but only a factor of two reduction in polarizability when the fields are orthogonal. These results may aid in stabilizing Rydberg resonances against varying dc fields present near surfaces, enabling advancement in the development of hybrid Rydberg atom - superconducting resonator quantum gates.
\end{abstract}

\maketitle

\section{Introduction}\label{Intro}

Neutral atom arrays are a promising  platform for quantum information processing thanks to their inherent scalability and the excellent coherence properties of the hyperfine ground states used to encode qubit states\cite{Saffman2019}. Strong Rydberg interactions between nearby atoms enable high-fidelity qubit entanglement. In addition to computation, atom arrays with Rydberg interactions are of interest in the development of quantum sensors, and for implementing quantum interfaces between different physical systems. One such example is the  use of Rydberg atoms coupled to microwave cavities as platforms for generating atom-superconducting qubit entanglement \cite{Petrosyan2009,Pritchard2014,Sarkany2015}, or transduction between optical and lower frequency microwave or mm wave fields\cite{Gard2017,Petrosyan2019,Covey2019, Kumar2023}. These experiments are enabled by the large transition dipole moments of microwave-frequency transitions between Rydberg states. Dipole allowed $nS-(n-1)P$ transitions have matrix elements  that scale as $\sim ea_0n^2$ \cite{Gallagher1994}, where $a_0$ is the Bohr radius, $e$ is the electronic charge, and $n$ is the principal quantum number. This is due to the large wavefunctions of Rydberg valence electrons, which scale in radius as $ \sim a_0 n^2$. Unfortunately, this large wavefunction also gives Rydberg atoms a very large dc polarizability, which scales as $\alpha \sim a_0^3 n^7$. This large polarizability means even modest stray electric fields cause large Stark-shifts. 

Slow drifts of background dc electric fields, often due to atom adsorbates on nearby surfaces, pose a persistent challenge for Rydberg atom based quantum systems. The nature  of adsorbate-caused fields, as well as other mechanisms leading to slowly varying fields near surfaces, has been studied in several works \cite{Tauschinsky2010,Kubler2010,Hattermann2012}. These drifts can shift Rydberg resonances away from the driving fields, and dc field gradients can broaden ground-Rydberg resonances, resulting in reduced coherence at small atom-surface distances\cite{Ocola2022}.

A variety of methods to measure, reduce, and stabilize adsorbate-caused electric field noise have been demonstrated \cite{Davtyan2018,
Hermann-Avigliano2014, Sedlacek2016, Thiele2014, Ocola2022}. However, in all of these examples large, often time-varying fields persist near the surfaces, with coherent Rydberg excitations closer than  $\sim90~\rm \mu m$ to any surface proving difficult. Microwave-atom entanglement schemes depend on coupling Rydberg atoms to evanescent resonator fields \cite{Hogan2012,Kaiser2022} or to fields extended past the resonator surface\cite{Beck2016}. In the former case, entering the strong-coupling regime requires atoms be placed within just a few $\rm\mu m$ of the resonator surface, in the latter example the requirement is relaxed to an atom-surface distance of $\sim15 ~\rm \mu m$.

A complementary approach to improving Rydberg state sensitivity in the face of dc electric field noise, is to reduce the Rydberg atom's dc polarizability via  an off-resonant microwave dressing field\cite{Mozley2005}. One such dressing technique, which reduced the differential dc polarizability between a pair of neighboring Rydberg  $S$  states has been demonstrated in \cite{Jones2013}. A dressing technique nulling the absolute and differential polarizabilities of circular Rydberg states was also proposed \cite{Ni2015}. Similarly, a dressing scheme where a two-tone dressing field is used to simultaneously reduce the absolute polarizability of a Rydberg $S$ state and a neighboring $P$ state has been proposed \cite{Booth2018}. The latter technique preserves not-only the Rydberg-Rydberg resonances, which are critical for cavity QED experiments, but also preserves the ground-Rydberg resonance necessary to perform the initial Rydberg excitation. Related dressing techniques have also been used to extend the coherence  of clock states separated by microwave transitions\cite{Sarkany2014}.

We demonstrate here reduction of the polarizability of $ 52P_{3/2}$ Rydberg states by up to a factor of 7 using a single frequency, elliptically polarized  dressing field.
The off-resonant microwave frequency at  $\omega_d = 2\pi \times 4780 ~\rm MHz$, admixes the $ 51D_{3/2}$ and $ 51D_{5/2}$ states into the $ 52P_{3/2}$ state. These $P$ and $D$ states have  opposite sign dc polarizabilities, resulting in a dressed $52P_{3/2}$ state with a reduced total polarizability.  The mixing angle of the dressed state is then varied by changing the amplitude of the dressing field, until a nulling condition is found, corresponding to  a dressing field amplitude where the dc polarizability of the dressed state is maximally reduced.  

The semi-classical system composed of the atom, the dc field and the dressing field is modeled numerically, guiding the experimental search for reductions of dc polarizability. The numerical model makes use of the Shirley-Floquet \cite{Shirley1965} technique to generate an effective Hamiltonian for the system, which is diagonalized to determine the spectrum of the dressed-atom system. Details of this model are presented in Section \ref{Model}. By observing how the modeled spectrum changes with different values of dc electric fields, the dc polarizability of dressed states are determined, and the theoretical nulling conditions are found. 

Section \ref{Experiment} presents the experimental apparatus and methods used to demonstrate polarizability reduction. The ground-Rydberg resonance is probed experimentally with  Cs atoms in a magneto-optical-trap (MOT). Atoms in the MOT which are excited to Rydberg states are rapidly photoionized by the MOT cooling and trapping beams, thus reducing the number of atoms in the MOT, which is measured as a reduction in flourescence on a camera. The response of the MOT depletion signal as the dc electric field in the vacuum chamber is changed is used to determine the polarizability of the atoms. The amplitude of the dressing field is then varied to determine the effects of the dressing field on the atom's polarizability. 
Results of the numerical modeling are presented in Sections \ref{ac stark} and \ref{Numerical}.
 The results of the experimental nulling search are then presented in Sec. \ref{Nulling}, followed by a summary of the results obtained in Sec. \ref{Conclusion}.

\section{Microwave Dressing Model} \label{Model}

\begin{figure}[!t]
    \centering
    \includegraphics[width=0.4\textwidth]{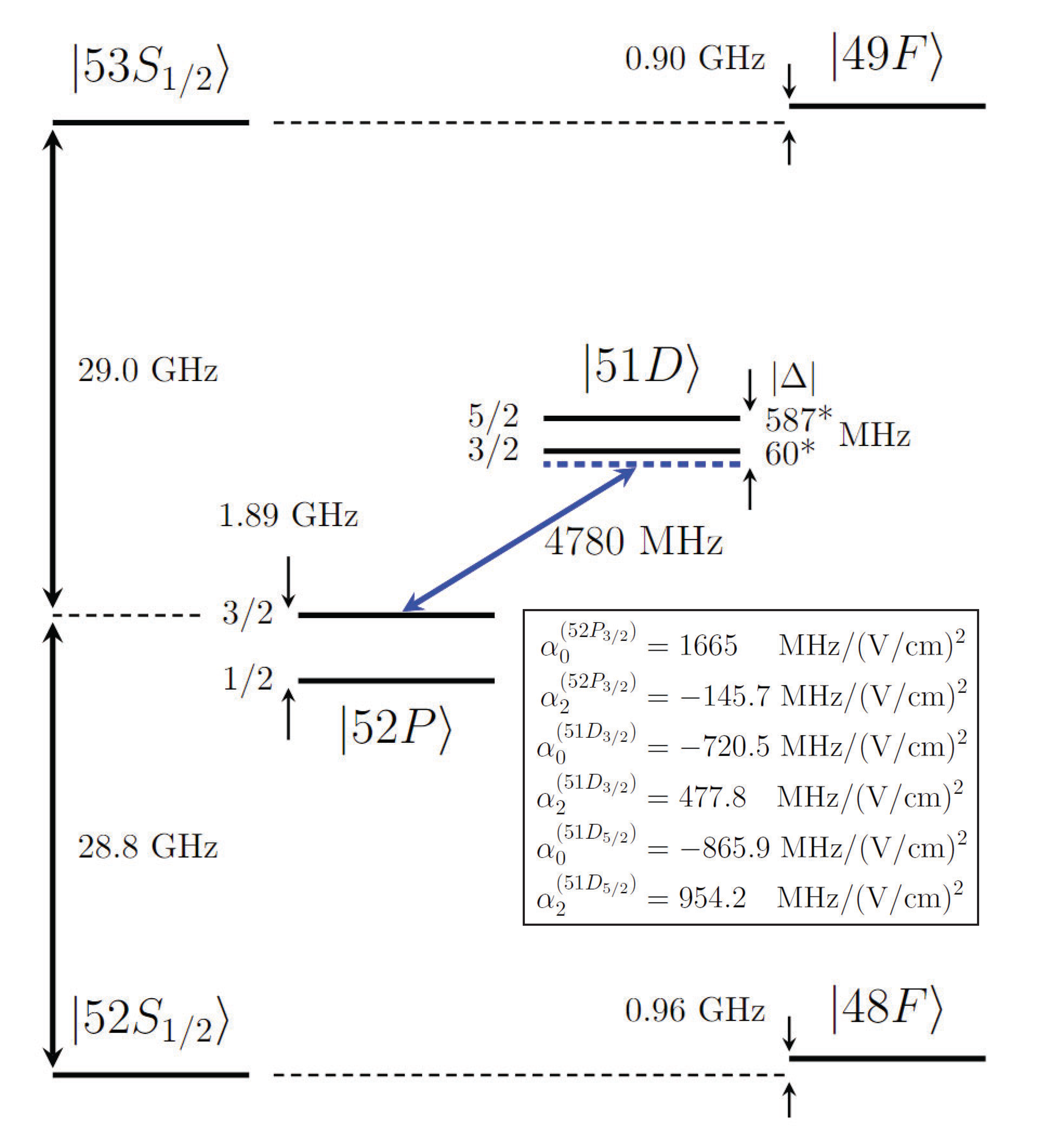}
    \caption{Energy level diagram of the dressed-atom system used to construct the model. The detunings between the dressing field and the $ 52P_{3/2}$-$ 51D_j$ Rydberg resonance are reported to the right of the respective $ 51D$ energy levels.\\
    $*$ Detunings modified from ARC values, based on measurement presented in Appendix \ref{Resonance}}
    \label{fig:SFGrotrian}
\end{figure}

The dressed-atom system is modeled by the Hamiltonian
\begin{equation} 
H(t) = H_0 + H_{\rm dc} + H_{\rm ac}(t)
\end{equation}
where $H_0$ is the unperturbed atomic Hamiltonian, $H_{\rm dc}$ models the effects of a dc electric field on the atom, and $H_{\rm ac}(t)$ models the effects of the time-dependent ac-dressing field on the atom. The matrix elements composing $H(t)$ were sourced from the Alkali Rydberg Calculator (ARC) \cite{Sibalic2016}. One modification was made, the $ 52P_{3/2} -  51D_{3/2}$ resonance frequency was directly measured to be  $4840~\rm MHz$ (see Appendix \ref{Resonance}) by microwave spectroscopy, starting from the optically excited $ 52P_{3/2}$ level. The measured resonance is $10 ~\rm MHz$ offset from the value provided by ARC. We attribute the difference to ac Stark shifts from the cooling and Rydberg lasers changing the atomic energy levels. The model was adjusted to reflect this change, by reducing the energy of the $ 52P_{3/2}$ level.   

$H_{\rm dc}$ is parameterized by the dc electric field strength $E_{\rm dc}$, and the orientation of the dc electric field
\begin{equation}
H_{\rm dc} = -E_{\rm dc}[  \sin(\theta)\cos(\phi){\hat{\bf x}}+\sin(\theta)\sin(\phi){\hat{\bf y}}+   \cos(\theta){\hat{\bf z}}  ] \cdot\mathbf{d}
\label{eq.Hdc_full}
\end{equation}
where $\theta, \phi$ are polar and azimuthal angles specifying the polarization of the dc field, and  $\mathbf{d}$ is the atomic dipole operator. Changing $\phi$ modifies the atomic spectrum in proportion to the ellipticity of the ac dressing field. In the system studied here, the ellipticity is small so the expected effects of changing $\phi$ are small, and have little effect on the analysis. Thus a simplified dc Stark Hamiltonian is used to investigate the dc response of the system, and it's anisotropic nature
\begin{equation}
H_{\rm dc} = -E_{\rm dc}[  \sin(\theta){\hat{\bf x}}+   \cos(\theta){\hat{\bf z}} ] \cdot\mathbf{d}.
\label{eq.Hdc}
\end{equation}
$H_{\rm ac}(t)$ is parameterized by the amplitude of the dressing field $E_{\rm ac}$, the frequency of the dressing field $\omega_{\rm d}$, and the small  ellipticity of the dressing field $\epsilon$
\begin{equation}
H_{\rm ac}(t) =-E_{\rm ac}[\sqrt{1-\epsilon}{\hat{\bf z}}+i\sqrt{\epsilon}{\hat{\bf y}}]\cdot\mathbf{d}\cos{(\omega_dt).}
\label{eq.Hac}
\end{equation}
The model Hamiltonian is constructed in a truncated basis consisting of Rydberg states that are energetically near the $ 52P_{3/2}$ level. The full model basis is shown in Figure \ref{fig:SFGrotrian}. This set of states, including the Zeeman sub-states,  was determined to be sufficient using convergence tests outlined in Appendix \ref{Convergence}.

The Shirley-Floquet (SF) technique is used to convert $H(t)$ into an effective, time-independent Hamiltonian $H_f$, which is then numerically diagonalized to compute the spectrum of the dressed-atom system\cite{Shirley1965,Booth2018}. $H_f$ is constructed from the Floquet theorem which states that for any time-periodic Hamiltonian $H(t)=H(t+T)$ there exists a solution to the Schr\"odinger equation $F(t)$, with the following form
\begin{eqnarray}
i\frac{dF(t)}{dt}&=&H(t)F(t),\nonumber\\
F(t) &=& \Phi(t)e^{-\imath Qt},\nonumber
\end{eqnarray}
where $\Phi(t)=\Phi(t+T)$ has the same periodicity as $H(t)$, and $Q$ is a diagonal matrix, whose entries $q_\alpha$ are the quasi-energies of the system
$$\braket{\alpha|Q|\beta} = q_\alpha\delta_{\alpha,\beta}.$$
$H_f$ is constructed by expanding $H(t)$ and $F(t)$ into their Fourier components, with the Fourier terms indexed by  
$m$
\begin{eqnarray}
H_{\alpha,\beta}(t) &=& \sum_m H^{(m)}_{\alpha,\beta}e^{\imath m\omega_dt}\nonumber\\
F_{\alpha,\beta}(t) &=& \sum_m F^{(m)}_{\alpha,\beta}e^{\imath m\omega_dt}e^{-iq_\beta t}\nonumber
\end{eqnarray}
where $H_{\alpha, \beta}(t)=\braket{\alpha|H(t)|\beta}$, and $F_{\alpha,\beta} = \braket{\alpha|F(t)|\beta}$. The Fourier expansions  of $H(t)$ and $F(t)$ are truncated with $-2\le m \le 2$ and substituted into the Schr\"odinger equation to form a set of recursion relations for $F^{(m)}_{\alpha,\beta}$, which can be re-written into a matrix-eigenvalue equation, of the form
$$H_f F_f = QF_f$$
where $H_f$ is the effective Shirley-Floquet Hamiltonian, and $F_f$ is the solution to the new, time-independent Schr\"odinger equation. $H_f$ is indexed in terms of the atomic quantum numbers (denoted by Greek letters), and a pseudo-quantum number that is the Fourier expansion index (denoted by Latin letters). In the case of $H(t)$ presented previously, the matrix-elements of $H_f$ are
\begin{multline*}
    H_{f,\alpha\beta}^{(m,n)} = \delta_{mn}(H_{0,\alpha \beta} + H_{\rm dc,\alpha \beta} - m\omega_d\delta_{\alpha\beta})+\\
    \frac{1}{2}(\delta_{n(m+1)}H_{\rm d,\alpha\beta}+\delta_{n(m-1)}H_{\rm d,\alpha\beta})
\end{multline*}

where
$$H_{\rm d} = -E_{\rm ac}[\sqrt{1-\epsilon}\mathbf{\hat{z}}+i\sqrt{\epsilon}\mathbf{\hat{y}}]\cdot\mathbf{d}$$
is the time-independent factor in $H_{\rm ac}$. The quasi-energies corresponding to states with $m=0$ are taken to be the approximate energies of the dressed Rydberg states.

The basis states of the system are the fine-structure Rydberg states shown in Fig. \ref{fig:SFGrotrian}, with the Fourier indices appended. For example $\ket{52P_{3/2},1/2;0}$ is the $\ket{52P_{3/2}, m_j=1/2}$ state, with Fourier index $m=0$. In total the basis is composed of 48 atomic states, with 240 atomic-Fourier states. The eigenstates of the dressed atom system are the dressed states. As the dressing field increases in amplitude, the dressed states become combinations of the basis states with high levels of mixing. A system for labeling the dressed states according to their inner-products with the basis states is developed. 

Small, non-zero values of $\epsilon$ cause mixing between the different magnetically sensitive $m_j$ states.   The $2j+1$ Zeeman states in a given Rydberg level are labeled with a band index $i$ that replaces the $m_j$ quantum number. The band index lies in the range $1\le i \le 2j+1$, ordered by the relative energies of the bands.
The label for a dressed state $\ket{\psi}$ is determined by computing the inner product of said dressed state with all basis states $c_{nlj m_j;m}$, then the values of the square of the inner products of each dressed state with each $m_j$ state in a Rydberg level are summed together
\begin{align} \label{eq.ovlp}
    c_{nlj m_j;m} &= \braket{nlj m_j;m|\psi},\\
    P_{nlj m_j;m} &= |c_{nlj m_j;m}|^2,\\
    P_{nlj;m} &= \sum_{m_j} P_{nlj m_j;m}.
\end{align}
Each dressed state is then labeled according to its largest $P_{nlj;m}$ value and assigned a band number, according to the energy ordering of the bands in that Rydberg level.

\section{Experiment}\label{Experiment}

\begin{figure}[!t]
    \centering
    \includegraphics[width=0.45\textwidth]{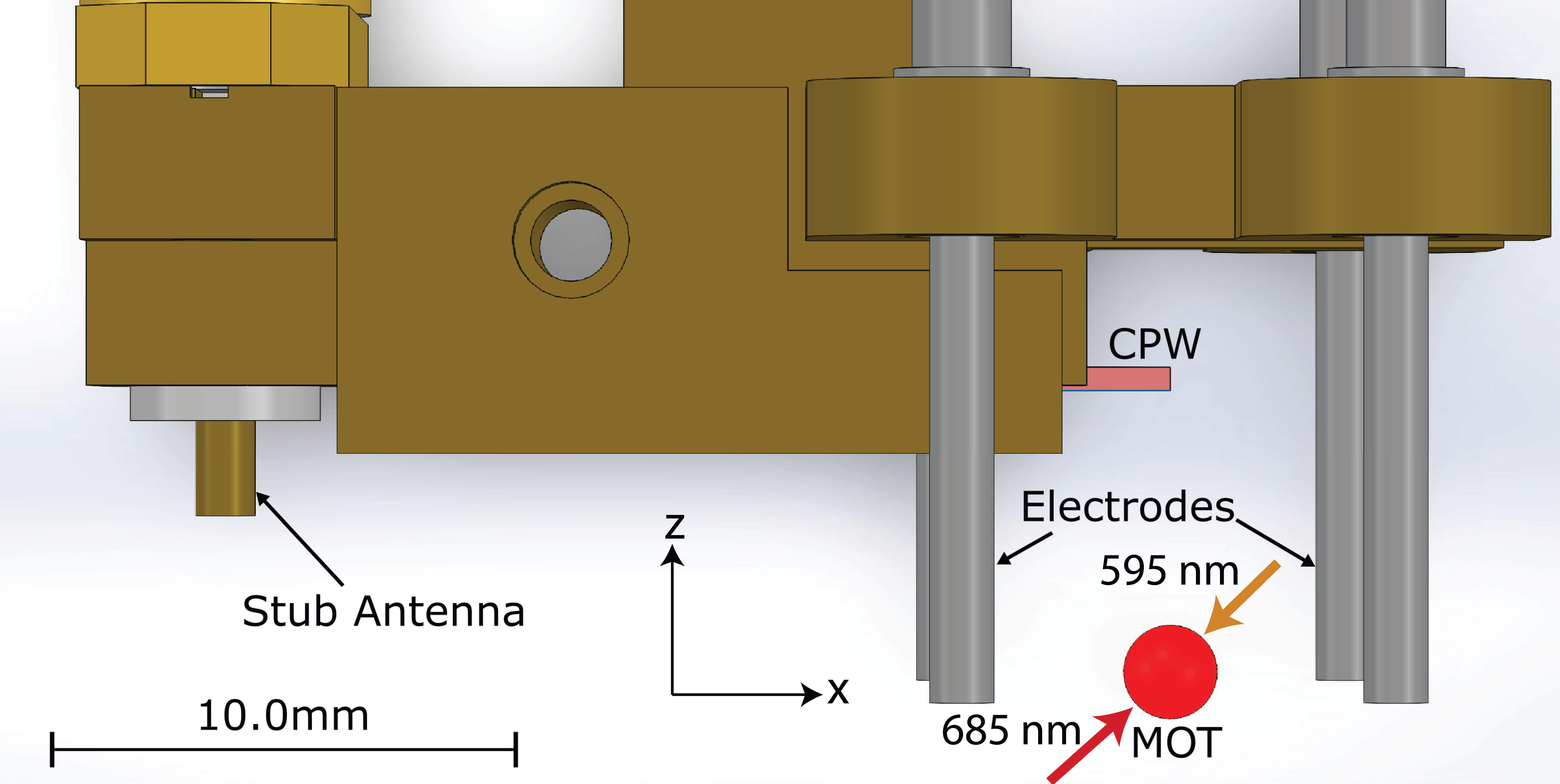}
    \caption{Schematic of the experimental apparatus inside the vacuum chamber. MOT (red circle) formed $\sim$ 5 mm below the CPW chip (pink rectangular structure), in the center of four electrodes (silver pins), which are used to vary the dc electric field at the MOT location. The dressing field is radiated by a stub microwave antenna (brass to the left), located  $\sim$ 20.5 mm away from the MOT. The Rydberg excitation beams counter-propagate in the $x-y$ plane and are polarized along $x$.  }
    \label{fig:Electrodes}
    \end{figure}

\subsection{ Experimental Apparatus} 
\label{Apparatus}

Measurements of the Rydberg polarizability were performed with neutral Cs atoms in a magneto-optical trap (MOT). The atoms are loaded into the MOT from atomic vapor,  laser cooled on the 
$\ket{ 6S_{1/2},F=4}$  - $\ket{6P_{3/2},F'=5}$ 
cycling transition and repumped on the 
$\ket{ 6S_{1/2},F=3}$  - $\ket{6P_{3/2},F'=4}$ 
transition. During the experiment cycle, the MOT cooling beams, MOT quadrupole field, and the Rydberg beams are left on continuously. The repumper is turned off at the start of each experiment cycle for a 10 ms long calibration phase, where the MOT beam powers are measured and feedback is applied, stabilizing the MOT beam powers. At the end of the experiment sequence, a 10 ms long exposure image is taken of the MOT, and the MOT intensity is inferred from the pixel counts inside a region of interest.

The MOT is formed inside of a 77 K cryostat that provides the intermediate temperature environment for atom-superconducting cavity QED experiments in which a single Cs atom is coupled to a superconducting coplanar-waveguide (CPW) resonator \cite{Pritchard2014, Beck2016}. The MOT is formed $\sim$ 5 mm below the CPW chip, which is mounted on a brass chip-mount shown in Fig. \ref{fig:Electrodes}. The chip-mount also holds four electrode pins and a stub microwave antenna. The electrodes (labeled pins 1-4) are used to vary the dc electric field at the MOT, while the stub antenna irradiates the atoms with the dressing field. The electrode pins are located on the corners of a
$25 \times 10~\rm mm$ rectangle, the MOT is located in the center of all four pins. The antenna is located about 20.5 mm from the MOT.

The microwave drive for the stub antenna is produced by a HP83623A signal generator, set to produce a CW tone at $\omega_d = 2\pi\times 4780~\rm MHz$. This tone is amplified (Mini-Circuits ZVE-3W-82+) and coupled to a directional coupler (Narda 25083) before being connected to the antenna feed-through on top of the cryostat. The power setting on the signal generator $P_{\rm s}$ is controlled to vary the amplitude of the dressing field. Inside the cryostat, the dressing field radiated by the stub antenna is scattered by the conducting surfaces making up the cryostat and chip-mount, before reaching the atoms. This scattering leads to an initially unknown polarization of the dressing field at the MOT. This unknown polarization is modeled by the inclusion of the $\epsilon$ parameter. The value of $\epsilon$, and the relationship between $P_{\rm s}$ and $E_{\rm ac}$ are determined by calibration procedures described in  Sec. \ref{ac stark}.

\begin{table}
    \centering
    \begin{tabular}{|c|c|c|c|}
        \hline
        Electrode & Pin 1 & All-1 & All-2 \\
        \hline
        Pin 1 & $V_{\rm s}$ & $V_{\rm s}$ & $V_{\rm s}$ \\
        Pin 2 & 0 & $0.421 V_{\rm s}$ & $V_{\rm s}$ \\
        Pin 3 & 0 & $0.6 V_{\rm s}$ & $V_{\rm s}$ \\
        Pin 4 & 0 & $0.6 V_{\rm s}$ & $V_{\rm s}$\\
        \hline
    \end{tabular}
    \caption{Voltage scaling on electrodes for the three configurations tested.}
    \label{tab:Pin Co}
\end{table}

The four electrode pins, in conjunction with the chip holder structure which is grounded,   can be used to produce electric fields in arbitrary directions. The electrode voltages are controlled by three bipolar digital-to-analog-converter boards (Analog Devices EVAL-AD5791SDZ-ND). Voltages on pins 1 and 2 are controlled independently, while pins 3 and 4 are set to the same voltage.

\begin{figure}[!t]
    \centering
    \includegraphics[width=0.4\textwidth]{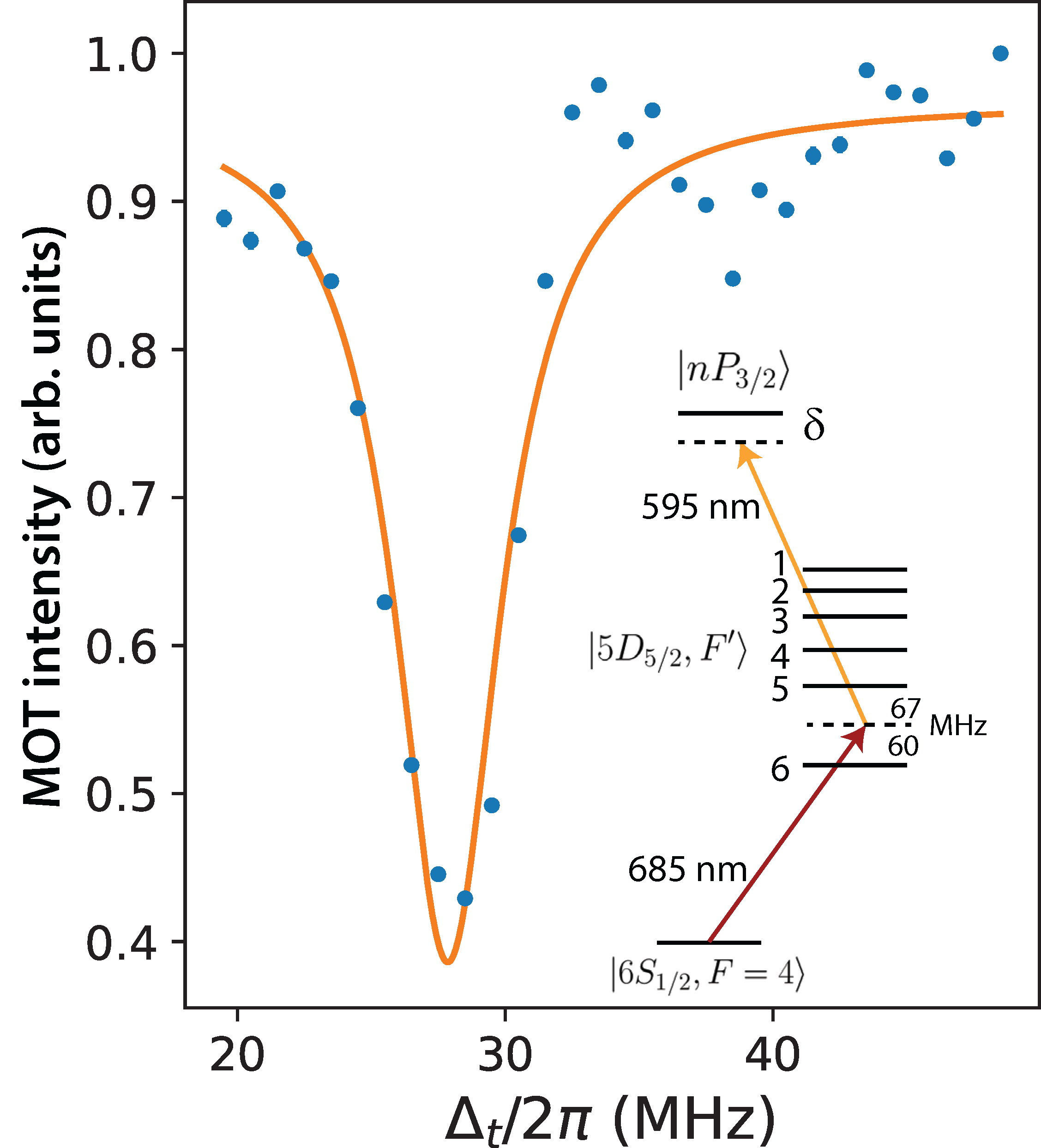}
    \caption{MOT depletion spectroscopy and Lorentzian fit. The Rydberg resonance occurs at $\mu_{\rm bare} = 27.85\pm0.015~\rm MHz$, with a line-width of $\gamma = 2.39\pm0.27~\rm MHz$. The inset shows the two-photon Rydberg excitation scheme used where the first photon transition is the $\ket{6S_{1/2}} \rightarrow \ket{5D_{5/2}}$ quadrupole transition at 685 nm. The second photon transition is the $ \ket{5D_{5/2}}\rightarrow \ket{52P_{3/2}}$ transition at 595 nm and $\delta$ is the two-photon ground-Rydberg detuning.    }
    \label{fig:Excitation}
\end{figure}

In order to test the dc Stark response of the atoms in different directions, thus probing the anisotropy of the dressed states' dc polarizability, three electrode configurations were chosen. Each electrode configuration had the voltages on the electrode pins scaled relative to the voltage on Pin 1, $V_{\rm s}$. The scaling constants for each electrode configuration are tabulated in Table \ref{tab:Pin Co}. The Pin 1 configuration only varied the voltage on electrode pin 1, while the All-1 pin configuration varied the voltages on all four electrodes, with scaling constants chosen such that the electric field strength produced by each electrode was roughly the same for  all values of $V_{\rm s}$.  The All-2 configuration was chosen to simply scale all electrode voltages to the same value. In order to determine the scaling values for the All-1 configuration, the un-dressed dc Stark response of the Rydberg state was measured for each electrode separately, and the scaling of the electric field strength was determined from the dc Stark response.

\subsection{Rydberg Spectroscopy}\label{rydberg}

\begin{figure} [!t]
    \centering
    \includegraphics[width=0.45\textwidth]{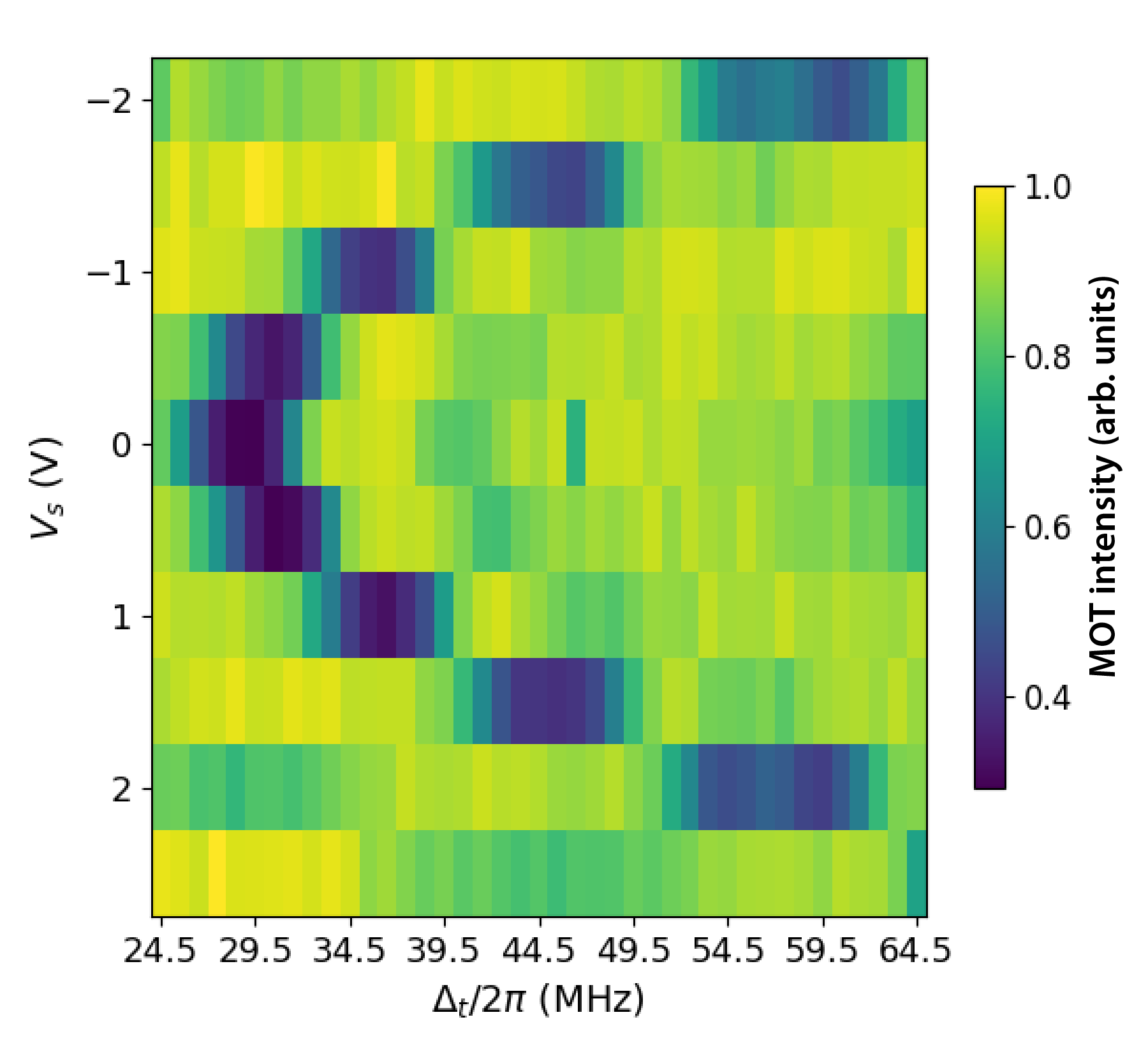}
    \caption{An example of MOT depletion spectroscopy used to measure the dc Stark response of the Rydberg states. The data were taken with the All-2 configuration of Table \ref{tab:Pin Co} with no dressing field.}
    \label{fig:All-2_Control}
\end{figure}

\begin{figure*}[!t]
    \centering    \includegraphics[width=0.9\textwidth]{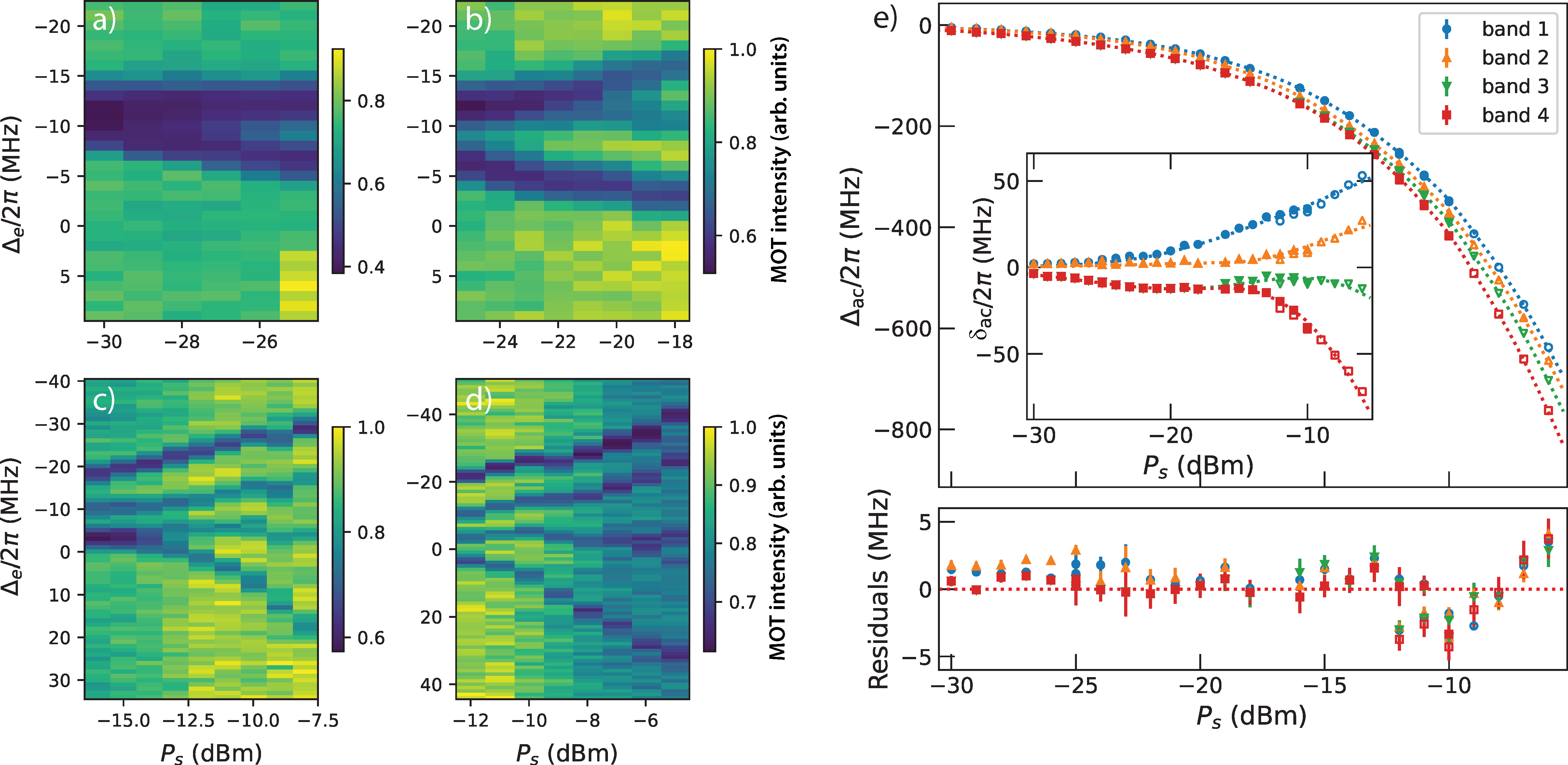} 
    \caption{ac Stark-shift spectroscopy calibration of the dressing field parameters. The measured resonances are compiled and compared to the dressing model, which uses two fit parameters to produce best-fit curves. \textbf{a)-d)} Experimental MOT depletion data. The plotted $\Delta_{\rm e}$ is the detuning from the Zeeman averaged ac Stark shift. \textbf{e)} Comparison of measured data with the theoretical shift using the estimated field amplitude and ellipticity.  The inset shows $\delta_{\rm ac}$ which is the same ac-Stark data with an approximate value of the mean of the four resonances subtracted out. The data from \textbf{d)} is excluded from the best-fit, and presented as un-filled shapes, as a verification of the model.}
    \label{fig:RF calibration}
\end{figure*}

Cs atoms are excited to Rydberg states using a  two-photon ladder excitation scheme with the first photon coupling to a quadrupole transition. As shown in Fig. \ref{fig:Excitation} the first photon  at 685 nm is detuned from the  $\ket{6S_{1/2},F=4} - \ket{5D_{5/2},F'=6}$ and $\ket{6S_{1/2},F=4} - \ket{5D_{5/2},F'=5}$ quadrupole transitions. The second photon  at 595 nm  is tuned so the sum frequency of the 685 nm and 595 nm fields is  near the two-photon ground-Rydberg resonance frequency
\begin{equation}
    \omega_{\rm 685}+\omega_{\rm 595} = \omega_{\rm R} - \omega_{\rm g} - \delta
\end{equation}
where $\omega_{\rm R}, \omega_{\rm g}$ are the Rydberg and ground state energies (divided by $\hbar$) and $\delta$ is the two photon detuning from ground-Rydberg resonance.  When $\delta=0$, the resonance condition is met, and Rydberg atoms are produced inside the MOT. 

The 685 nm and 595 nm fields are produced by two counter-propagating laser beams, focused to waists ($1/e^2$ intensity radii) of  14 $\rm \mu m$ and 5 $\rm \mu m$ respectively. The two Rydberg beams are made to overlap inside the MOT. 
The 685 nm light is generated by a M2-Solstis Ti:Sa laser. 
The 595 nm light is generated by second-harmonic-generation  using a PPKTP crystal inside a resonant bowtie cavity, pumped by an 1190 nm Toptica diode laser. The 685 nm and 1190 nm lasers are frequency stabilized to Ultra-Low-Expansion (ULE) reference cavities using the Pound-Drever-Hall(PDH) \cite{Drever1983} technique. The 1190 nm light is detuned from a ULE resonance by a value $\Delta_{\rm t}$, which is tuned to probe the ground-Rydberg resonance. 
The shift $\Delta_{\rm t}$ is controlled with a broadband fiber electro-optic-modulator (EOM) between the 1190 nm laser and the ULE cavity. The ULE cavities are housed in temperature stabilized vacuum cans  resulting in absolute frequency drifts of approximately 10 kHz/day.  The EOM is driven by two tones, one at $\Delta_{\rm t}$, and one at the PDH sideband frequency.  The sideband at $+\Delta_{\rm t}$ is  locked to the ULE cavity, thus when $\Delta_{\rm t}$ is changed slowly, the +1 side-band remains locked, and the 1190 nm laser is tuned with an opposite sign to the change. Thus when  $\Delta_t$ increases (decreases),  the frequency of the 595 nm light  decreases (increases) by twice that value due to the frequency doubling. The sideband shift  $\Delta_{\rm t}$ can take on values between 15 MHz and 1.5 GHz.

MOT depletion spectroscopy has been used to measure the ground-Rydberg resonances and how they respond to the dc and dressing fields. MOT depletion occurs when atoms inside a MOT are excited to Rydberg states, these Rydberg atoms are rapidly photo-ionized by the MOT cooling and trapping light, reducing the number of atoms inside the MOT. The reduced atom number is observed as a reduction in MOT flourescence, which is detected on an EMCCD camera (Andor Luca R 604).  At each value of the scanned  experimental parameter 50 images are taken,  and are then averaged.  

A typical MOT depletion spectroscopy curve is shown in Fig. \ref{fig:Excitation}. The depletion signal was used to find the ground-Rydberg resonance in the absence of dc or dressing fields, which occurred at  $\Delta_{\rm t}/2\pi=\mu_{\rm bare} =  27.85 \pm 0.015 ~\rm MHz$ relative to a chosen ULE resonance. In order to measure the dc Stark response of the $ 52P_{3/2}$ states, the voltages on the electrode pins are changed and the Rydberg spectroscopy is repeated at each value of $V_{\rm s}$. The dc Stark response of an undressed Rydberg state can be seen in Fig. \ref{fig:All-2_Control}, where the electrodes are in the All-2 configuration. We note that the resonance shift $\Delta_{\rm t}$ is a minimum when $V_{\rm s}=0$. This corresponds to a maximum of the Rydberg energy, and since the $52P_{3/2}$ polarizability is positive we infer that the background electric field in the chamber is negligibly small.

\section{Results} \label{Results}

\subsection{ac Stark-Shift} \label{ac stark}

The amplitude and ellipticity of the dressing field were calibrated by measuring the ac Stark-shift on all four $ 52P_{3/2}, m_j$ states as a function of dressing field power and comparing with the numerical model.  Four experimental scans  were taken with $P_{\rm s}$ varied between -30 dBm and -5 dBm. At each value of $P_{\rm s}$ the MOT depletion spectroscopy resonances were fit to a function consisting of multiple Lorentzian lineshapes, where the line-centers are the ground-Rydberg resonances. The centers of these lineshapes were labeled $\mu_{i}$, where  $i$ is the state's band index. The ac Stark shift was computed by comparing the observed resonances to $\mu_{\rm bare}$
\begin{align} \label{eq:dac_def}
    \Delta_{{\rm ac},i} &= 2(\mu_{i}-\mu_{\rm bare})
\end{align}
The factor of two is due to the frequency doubling of $\Delta_{\rm t}$. 

The data from the ac Stark experiments are presented in Fig. \ref{fig:RF calibration}a)-d). The values of $\Delta_{\rm t}$ sampled in the experimental data are referenced to an estimate of the average ac-Stark shifted resonance of all four bands $\mu_e(P_s)$
\begin{equation} \label{eq.Deltae}
    \Delta_{{\rm e}} = \Delta_{{\rm t}}-\mu_e(P_s).
\end{equation}
Details on the computed values of $\mu_e(P_s)$ are in Appendix \ref{CoG}.

The experimental data was then compared to numerical results produced using the SF model. The numerical data was generated by varying $E_{\rm ac}$ between $0 - 90 ~\rm{V/m}$, at each amplitude value $H_f$ was diagonalized, the $ 52P_{3/2}$ bands were identified and their ac Stark-shift was computed.
In the fit of the numerical data to the experiment, two fit parameters were used $E_0$ and $\epsilon$ with  $E_0$ the  dressing field amplitude when $P_{\rm s}=0 ~\rm dBm$, and $\epsilon$  the ellipticity in the field, as specified in Eq. (\ref{eq.Hac}).

Residuals between the experiment and numerical fit were produced for each set of $E_{\rm ac}, \epsilon$ values, by first re-scaling the field amplitude in the numerical data according to
$P_{\rm s} = 20 \log_{10}\left(E_{\rm ac}/E_0\right).$ 
The numerical data was then interpolated to compute the residuals at the experimentally sampled values of $P_{\rm s}$.
For each set of $E_0,~\epsilon$ values, the $\chi^2$ of the fit was computed. The values of $E_0,~\epsilon$ that minimized the $\chi^2$ were taken to be the best-fit parameters resulting in $E_0=167 ~\rm V/m$ and
$\epsilon=0.012$. The best-fit curves are shown in Fig. \ref{fig:RF calibration}e. In the inset of the Figure, values for $\mu_e(P_s)$, newly computed using the experimental ac-Stark shift data,  are subtracted out of the experimental and numerical data, this demonstrates the fine details of the ac Stark shifted spectrum between the different bands in the $52P_{3/2}$ level.

The ac Stark shift data gathered in the first three experiments (Fig. \ref{fig:RF calibration} a-c), were used in the fit calibration of the experimental data. The data from the fourth experiment (Fig. \ref{fig:RF calibration} d) was excluded from the fits shown in Fig. \ref{fig:RF calibration}e , and is presented alongside the rest of the dataset as a way of validating the fit at high dressing powers. For that reason the high power ac Stark data are presented in Fig. \ref{fig:RF calibration}e as open shapes.

In Fig. \ref{fig:RF calibration}e, the fourth experiment redundantly samples values of $P_{\rm s}$ from $-12 ~\rm dBm$ to $-10 ~\rm dBm$, there is $\sim 4$ MHz discrepancy between those values sampled in the first three experiments, and the fourth experiment. This discrepancy is indicative of the systematic uncertainty present in the experimental procedure. These experiments were run with a four day gap between them, thus uncontrolled fluctuations in the background dc electric field strength, which may have accumulated in that time, are expected to be the most prominent contribution to the discrepancy.

\begin{figure}[!t]
    \centering
    \includegraphics[width=0.5\textwidth]{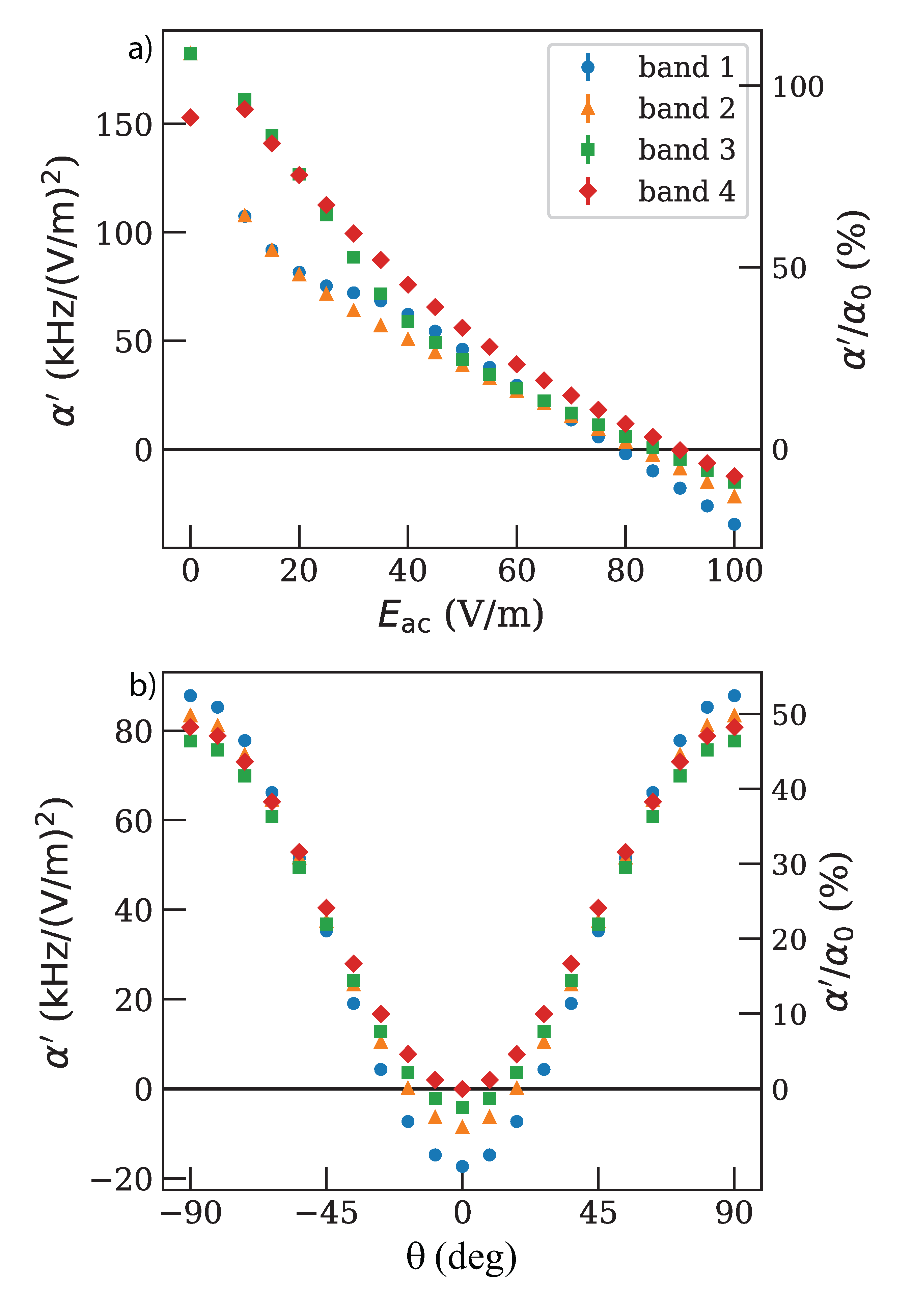}
    \caption{Numerical Estimates of dc polarizability reduction produced by varying the dc field from 0-100 V/m and fitting the Stark response to a fourth order polynomial. The dressed dc polarizability ($\alpha'$) is compared to the (scalar) un-dressed polarizability ($\alpha_0$). \textbf{a)} dc Polarizability vs $E_{\rm ac}$. From these data we find the zero-crossing for band 4 at $89.6 ~\rm V/m$. \textbf{b)} dc Polarizability with $E_{\rm ac}=89.6$ V/m as $\theta$ is varied. The anisotropy of the polarizability reduction is clear from this plot, when the dc and dressing fields are orthogonal the dressed state retains $\rm \sim 50\%$ of it's undressed polarizability.}
    \label{fig:Numerical Polarizability}
\end{figure}

\subsection{Numerical Nulling} \label{Numerical}

\begin{figure*}[t]
    \centering
    \includegraphics[width=0.90\textwidth]{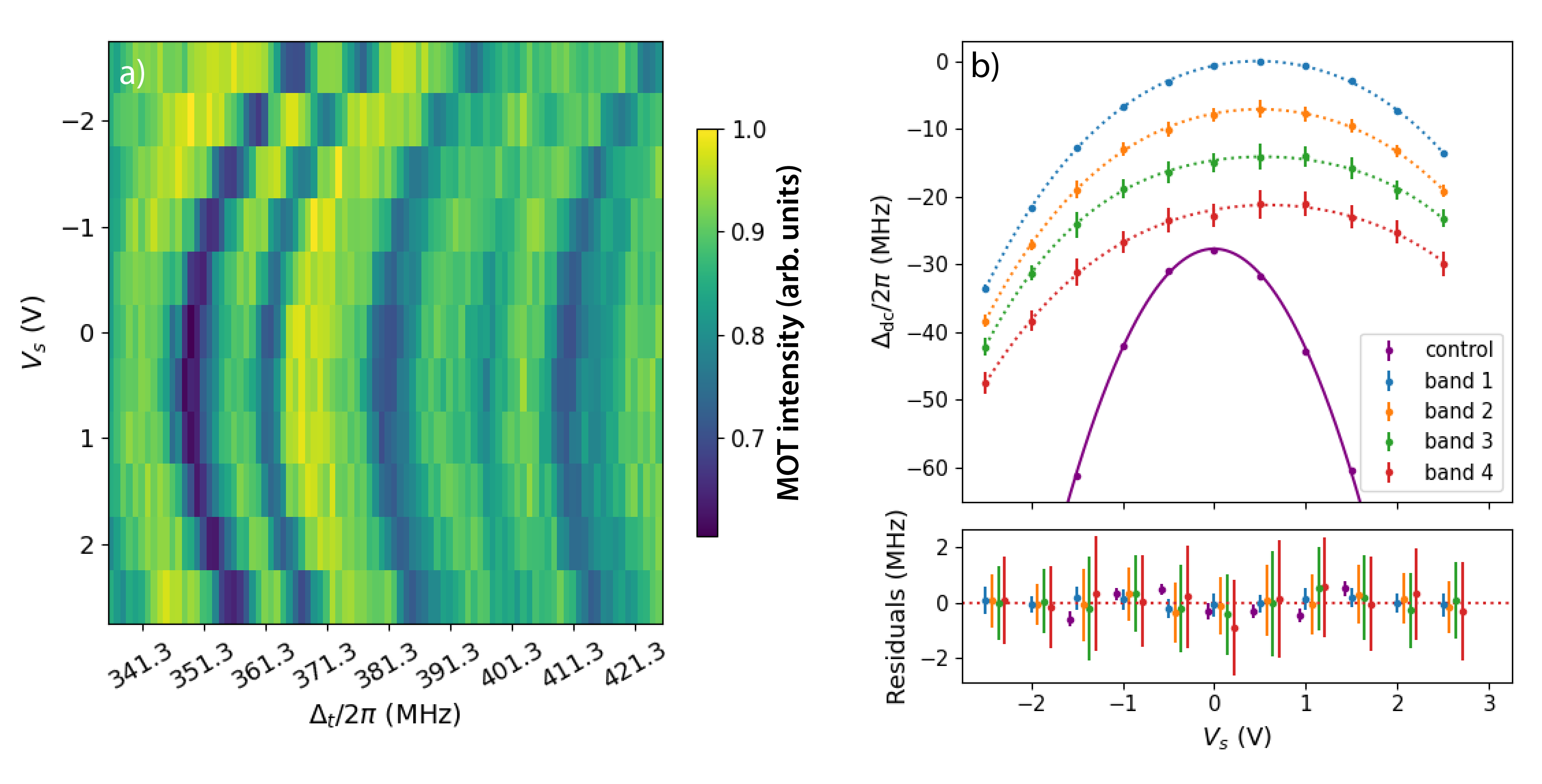}
    \caption{Polarizability reduction demonstrated using the All-2 pin configuration. \textbf{a)}: MOT depletion dc spectroscopy taken  in the All-2 Configuration with the dressing field on. Four bands are clearly visible, along with an Autler-Townes band near 400 MHz (see Sec. \ref{AT}), this band is excluded from the dc Stark analysis. \textbf{b)}: The measured dc Stark-shifts in the control and dressed experiments using the All-2 configuration. The $x$-axis is the $V_{\rm s}$ parameter described in Table \ref{tab:Pin Co}. The control data are fit to a 2nd-order polynomial, while the dressed data are fit to a 4th-order polynomial. The four bands and control data are offset by 7 MHz on the y-axis for clarity.}
    \label{fig:All-2 dressing} 
    
\end{figure*}

\begin{table}[t]
\begin{tabular}{|c|c|c|}
    \hline
    Band & Nulling Field Amplitude (V/m) & Nulling Power (dBm) \\
    \hline
    1 & 78.6 & -6.5\\
    2 & 82.6 & -6.1\\
    3 & 85.6 & -5.8\\
    4 & 89.6 & -5.4 \\
    \hline
    \end{tabular}
    \caption{Numerically computed polarizability nulling conditions for all four bands, and the $P_{\rm s}$ values which null polarizability in each band, based on the calibration in Sec. \ref{ac stark}.}
    \label{tab:nulling conditions}
\end{table}

The calibrated SF model is used to find the dressing field amplitude that nulls the dc polarizability of the dressed $ 52P_{3/2}$ Rydberg states with model parameters $\omega_d = 2\pi\times4780$ MHz, $\epsilon=0.012$, and $\theta=0$. The polarizability of any dressed state is determined by fitting the dc Stark response of each state to a fourth-order polynomial \cite{Davydkin1986}
\begin{equation} 
\Delta U/h = -\frac{1}{2}\alpha'E_{\rm dc}^2-\frac{1}{4!}\beta'E_{\rm dc}^4\end{equation} 
Where $\alpha'$ is the effective polarizability of the dressed state, and $\beta'$ is the effective hyper-polarizability of the dressed state. The dc response is evaluated for values of $E_{\rm dc}$ between 0-15 V/m.

The nulling condition is found by computing $\alpha'$ for each band for values of $E_{\rm ac}$ between 0-100 V/m. The computed values of $\alpha'$ are plotted in Fig. \ref{fig:Numerical Polarizability}. The resulting values of $\alpha'$ are then interpolated and the zero-crossing for each band, corresponding to it's nulling condition, is found. The nulling conditions for each band are tabulated in Table \ref{tab:nulling conditions}, along with the estimated values of $P_{\rm s}$.

The effects of misalignment between the dc and dressing fields are investigated by setting the dressing field amplitude to the nulling condition for band 4, $E_{\rm ac} = 89.6 ~\rm V/m$, then $\theta$ is scanned from $-90^\circ$ to $90^\circ$. The result of this scan are plotted in Fig. \ref{fig:Numerical Polarizability}. The anisotropy of the dressed-atom system is evident, with the dressed-atom retaining about 50\% of its un-dressed polarizability when the fields are orthogonal. It's also possible to null the polarizability of a dressed state, when the dressing field amplitude is higher than the nulling condition at $\theta=0$, by introducing a small angle between the dressing and dc fields, as can be seen in bands 1-3.

\subsection{Experimental Polarizability Reduction} \label{Nulling}

\begin{table*}[t]
\begin{tabular}{|c|c|c|c|c|c|c|}
    \hline
     & \multicolumn{2}{|c|}{Pin 1} & \multicolumn{2}{|c|}{All-1} & \multicolumn{2}{|c|}{All-2} \\
     \hline
    Band & $\alpha'$ ($\rm MHz/V^2$) & $\alpha'/\alpha'_{\rm c}$ (\%) & $\alpha'$ ($\rm MHz/V^2$) & $\alpha'/\alpha'_{\rm c}$ (\%) & $\alpha'$ ($\rm MHz/V^2$) & $\alpha'/\alpha'_{\rm c}$ (\%)\\
    \hline
    Control &  $7.14 \pm 0.04$ & $100$ & $10.3 \pm 0.008$ & $100$ & $29.4\pm0.54$ & $100$\\
    Dressed 1 & $3.54\pm0.06$ & $50\pm1$ & $2.50\pm0.22$ & $24\pm2$ & $5.82\pm0.12$& $20\pm1$ \\
    Dressed 2 & $3.42\pm0.18$ & $48\pm3$ & $2.24\pm0.24$ & $22\pm3$ & $5.10\pm0.18$& $17\pm1$ \\
    Dressed 3 & $3.42\pm0.28$ & $48\pm4$ & $2.30\pm0.24$ & $22\pm3$ & $3.70\pm0.24$& $13\pm1$ \\
    Dressed 4 & $4.02\pm0.28$ & $56\pm4$ & $1.74\pm0.22$ & $17\pm2$ & $4.00\pm0.32$& $14\pm1$ \\
    \hline
    \end{tabular}
    \caption{Polarizabilities in all three tested configurations. The fit values for $\alpha'$ and the remaining percentage of the polarizability are reported}
    \label{tab:pol}
\end{table*}

Tests of the polarizability reduction were performed for all three electrode configurations in two steps: \textbf{1.} a control experiment with no dressing field present is used to determine the magnitude of the dc electric field produced, per volt, in a given pin configuration and \textbf{2.} a dressed experiment where the dressing field was turned on, with $ P_s=-6 ~\rm dBm$. This value was chosen as it's near the nulling condition for bands 2 and 3. In both experiments $V_{\rm s}$ was varied from a negative to a positive voltage, with the voltage ranges chosen to produce effective fits to the dc Stark data.

In the control experiments the dressing field is turned off by programming $ P_{\rm s}=-100 ~\rm dBm$. Then MOT depletion spectroscopy was performed for each value of $V_{\rm s}$, and the depletion data was fit to a single Lorentzian at each voltage. The dc Stark-shift was then computed for each value of $V_{\rm s}$ as
\begin{align} \label{eq.dc_stark}
  \Delta_{\rm dc} = 2(\mu-\mu_{\rm min}) 
\end{align}
Where $\mu$ is the line center at each value of $V_{\rm s}$, and $\mu_{\rm min}$ is the smallest value of $\mu$ in the parabola. The electric field strength per  volt is then calibrated by fitting the Stark-shift data to the second order polynomial
\begin{align}\label{eq.exp_2fit}
\Delta_{\rm dc}\left(V_{\rm s}\right) = -\frac{1}{2}\alpha'_{\rm c} \left(V_{\rm s}-V_0\right)^2
\end{align}
Where $\alpha'_{\rm c}$ is the effective voltage-polarizability of the state in the given electrode configuration, in units of $~\rm MHz/V^2$, and $V_0$ is the center of the parabola, accounting for the effects of any stray dc electric fields present. The voltage polarizability of the Rydberg state in a given electrode configuration can be compared to the computed value of the scalar polarizability of the $52P_{3/2}$ state, $\alpha_0 = 1664.98 ~\rm MHz/(V/cm)^2$, to extract the conversion between $E_{\rm dc}$ and $V_{\rm s}$.

For the dressed experiments, the MOT depletion spectroscopy was performed at each value of $V_{\rm s}$, and the depletion data was fit to a function consisting of four Lorentzian features. The dc Stark-shift was then computed for each of the four resonances
\begin{align} \label{eq.dc_stark_dressed}
    \Delta_{{\rm dc},i} &= 2(\mu_i-\mu_{i,\rm min})
\end{align}
and the dc response of each resonance was fit to the fourth-order polynomial
\begin{equation}
\Delta_{\rm dc}(V_s) = -\frac{1}{2}\alpha' (V_s-V_0)^2-\frac{1}{4!}\beta(V_s-V_0)^4\end{equation}
where $\alpha'$ is the dressed state's effective voltage-polarizability, and $\beta'$ is the dressed state's effective voltage-hyper-polarizability. In the Pin 1 configuration only a second order fit was required to match the observed data, so the fourth-order term was omitted. This is likely due to a smaller hyper-polarizability for dressed states when the electric field is orthogonal to the dressing field, as can be seen in Appendix \ref{4O}. In the All-1 and All-2 cases the fourth-order term was necessary to adequately fit to the experimental data.

The measured control and dressed effective polarizabilities are tabulated in Table \ref{tab:pol}. The All-2 electrode configuration had the greatest reduction in polarizability, with band 3 demonstrating more than 7 times reduction in measured polarizability. The All-2 data is shown in Fig. \ref{fig:All-2 dressing}, where the fit to the control data is represented with a solid line, while the fits to the dressed data are represented with dotted lines.

In all three electrode configurations, there are reductions in the polarizability of the dressed states. The smallest reduction was observed in the Pin 1 configuration, where about 50\% of the undressed atoms polarizability is retained, as is predicted by the model. The different response of the dressed atoms in these different configurations is indicative of the anisotropy of the dressed-atom system, as each configuration generated electric fields in different directions.

\section{Conclusion} \label{Conclusion}

In summary, we have demonstrated a reduction of the dc polarizability of $ 52P_{3/2}$ Rydberg states in Cs using an off-resonant microwave dressing field. The anisotropy predicted by numerical modeling was   probed and demonstrated by varying the direction of the dc field when testing the polarizability of the dressed states. With a dressing field amplitude of $E_{\rm ac}\sim 87 ~\rm V/m$, the polarizability of one $52P_{3/2}$ state was reduced by 7 times in one direction, but only about 2 times in another. A combination of numerical and experimental data were used to determine the ellipticity of the dressing field, and the necessary dressing field power to null the dc polarizability in some axis. This technique can be a powerful tool when improving the fidelity of ground-Rydberg excitations inside systems with moderately noisy electric fields. Though the significantly anisotropic nature of the dressed state polarizability limits the utility of this dressing scheme to systems where the direction of the electric field noise is known, such as systems where atoms are placed near conducting surfaces and the boundary conditions set the dc fields normal to the conducting surfaces. 
The technique used to model this system can be readily applied to different Rydberg states in Cs or other atomic species, and different dressing frequencies (see Appendix \ref{Alternates}). Thus allowing a range of control options for achieving  low dc polarizability. 

Another feature of this technique is that the dc-Stark response is only reduced in small dc fields. Above field strengths of about 20 V/m the hyper-polarizability produces much larger Stark shifts in the dressed states than in the un-dressed states. This poses a significant challenge for near-surface experiments, but techniques from \cite{Davtyan2018,
Hermann-Avigliano2014, Sedlacek2016, Thiele2014, Ocola2022} can be used in conjunction with dressing to improve the stability of the Rydberg resonance. When approaching atom-surface distances $< 100 ~\rm \mu m$, it may become necessary to use multi-frequency dressing to reduce the higher order polarizabilities, as proposed in \cite{Ni2015, Booth2018}.

\begin{acknowledgments}
Research supported by NSF Grant No. 2016136 for the
Quantum Leap Challenge Institute center Hybrid Quantum Architectures and Networks. JCB was supported by a National Science Foundation Graduate Research Fellowship Program under Grant No. DGE-1747503. We appreciate technical help in developing the apparatus from Jonathan Pritchard,  Danny Wendt, Sebastian Malewicz, and Bradley Nordin. 
Any opinions, findings, and conclusions or recommendations expressed in this material are those of the authors and do not necessarily reflect the views of the National Science Foundation.
\end{acknowledgments}

\bibliography{atomic.bib,saffman_refs.bib,rydberg.bib,qc_refs.bib,optics.bib,QED.bib}

\clearpage
\appendix

\section{Convergence Tests} \label{Convergence}
\begin{figure}[!t]
    \centering
    \includegraphics[width=0.5\textwidth]{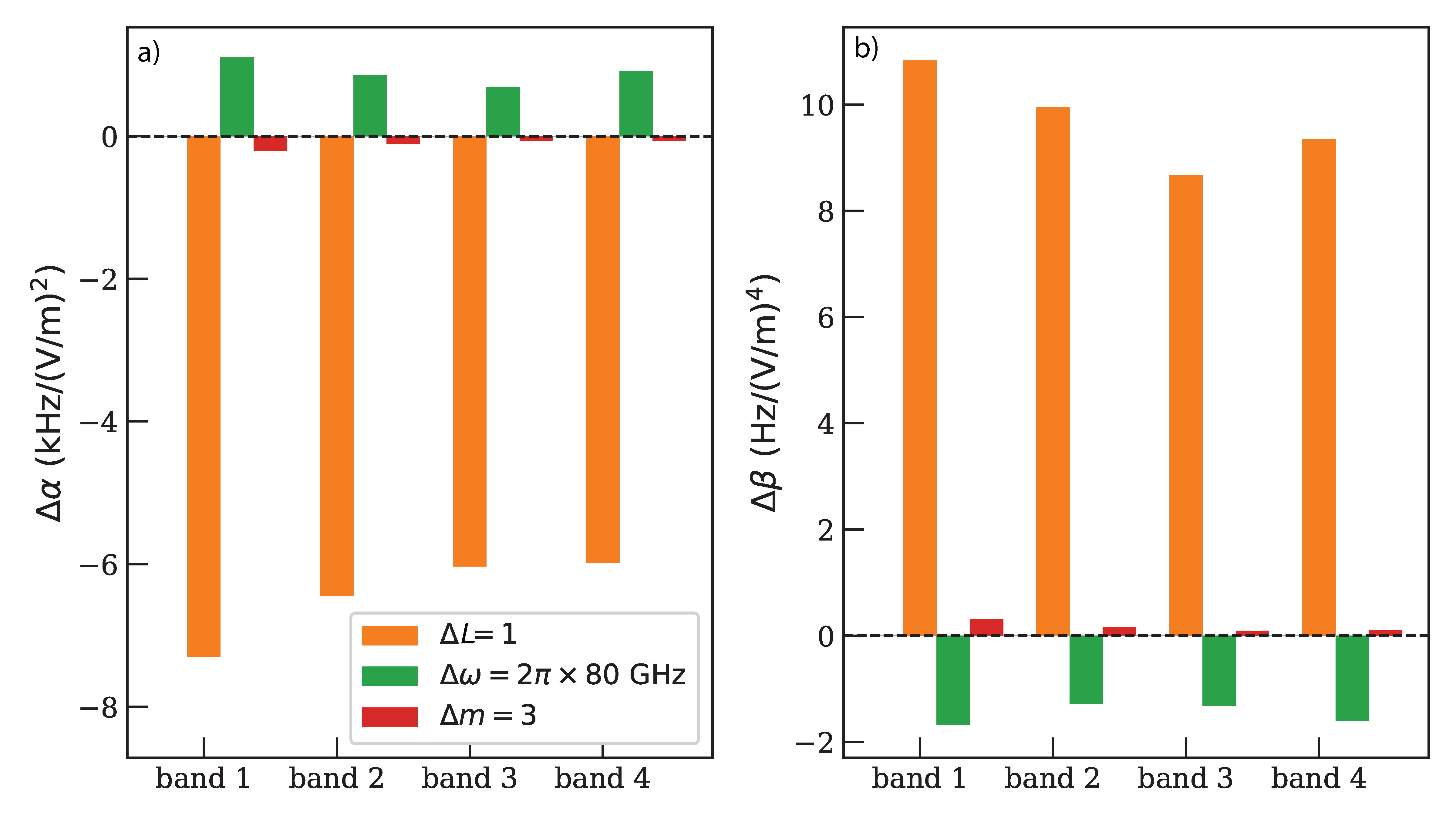}
    \caption{Results of convergence tests of parameters determining the size of the modeled Hilbert space. \textbf{a)} Change in polarizability and \textbf{b)} Change in hyper-polarizabiliy are compared to the model used in the main text. These values were computed at the band 4 nulling condition, $E_{\rm ac}=89.6~\rm V/m$, for  Hilbert spaces which differ in size from that of the main text. }
    \label{fig:Convergence}
\end{figure}

\begin{table}[h]
\begin{tabular}{|c|c|c|c|c|}
    \hline
     Parameter & Main Text & $\Delta L = 1$ & $\Delta \omega / 2\pi$ & $\Delta m = 3$ \\
    \hline
    $\Delta \omega/2\pi~\rm (GHz)$ &  40  & 40 & 80 & 40\\
    $\Delta L$ & 2 & 1 & 2 & 2 \\
    $\Delta m$ & 2 & 2 & 2 & 3 \\
    \hline
    \end{tabular}
    \caption{List of parameter sets used to vary the Hilbert space dimensions for convergence tests.}
    \label{tab:Convergence}
\end{table}

The modeled Hilbert space with the atomic levels indicated in Fig.\ref{fig:SFGrotrian} was constructed systematically, based on the target level $ 52P_{3/2}$ and three computational parameters:
\begin{itemize}
    \item \textbf{$\Delta \omega$} : the largest energy difference between included  Rydberg states and the target state.
    \item \textbf{$\Delta L$} : the largest difference in $L$ quantum number between included Rydberg states, and the target state.
    \item \textbf{$\Delta m$} : the number of Fourier indices of included states before truncation (positive and negative)
\end{itemize}

Three different cases were compared to the computational parameters of the model presented in the main text, these changes are tabulated in Table \ref{tab:Convergence}.

Convergence was tested by computing the polarizability and hyper-polarizability of the four bands in $52P_{3/2}$ in all four configurations. This was done at the band 4 dressing condition, $E_{\rm ac} = 89.6 ~\rm V/m$ and $\epsilon=0.012$. These results are then compared to the configuration in the main text in Fig. \ref{fig:Convergence}.

The $\Delta L = 1$ case excludes the nearby $nF$ levels, and demonstrates the importance of their inclusion for the model's accuracy. The $\Delta \omega = 2\pi \times 80 ~\rm GHz$ case expands the Hilbert space by including the nearby $51P$, $53P$, $50D$ and $52D$ levels. The $\Delta m = 3$ case expands the number of Fourier components included in the truncated $H_f$.

While the inclusion of the nearby $F$ levels provides a necessary contribution, the nearest $P$ and $D$ levels only provide a correction $\sim 0.01\alpha_0$, where $\alpha_0$ is the undressed scalar polarizability of the $52P_{3/2}$ state. Expanding the number of included Fourier components provided a very small correction.

\section{Hyper-polarizability} \label{4O}

\begin{figure}[!t]
    \centering
    \includegraphics[width=0.45\textwidth]{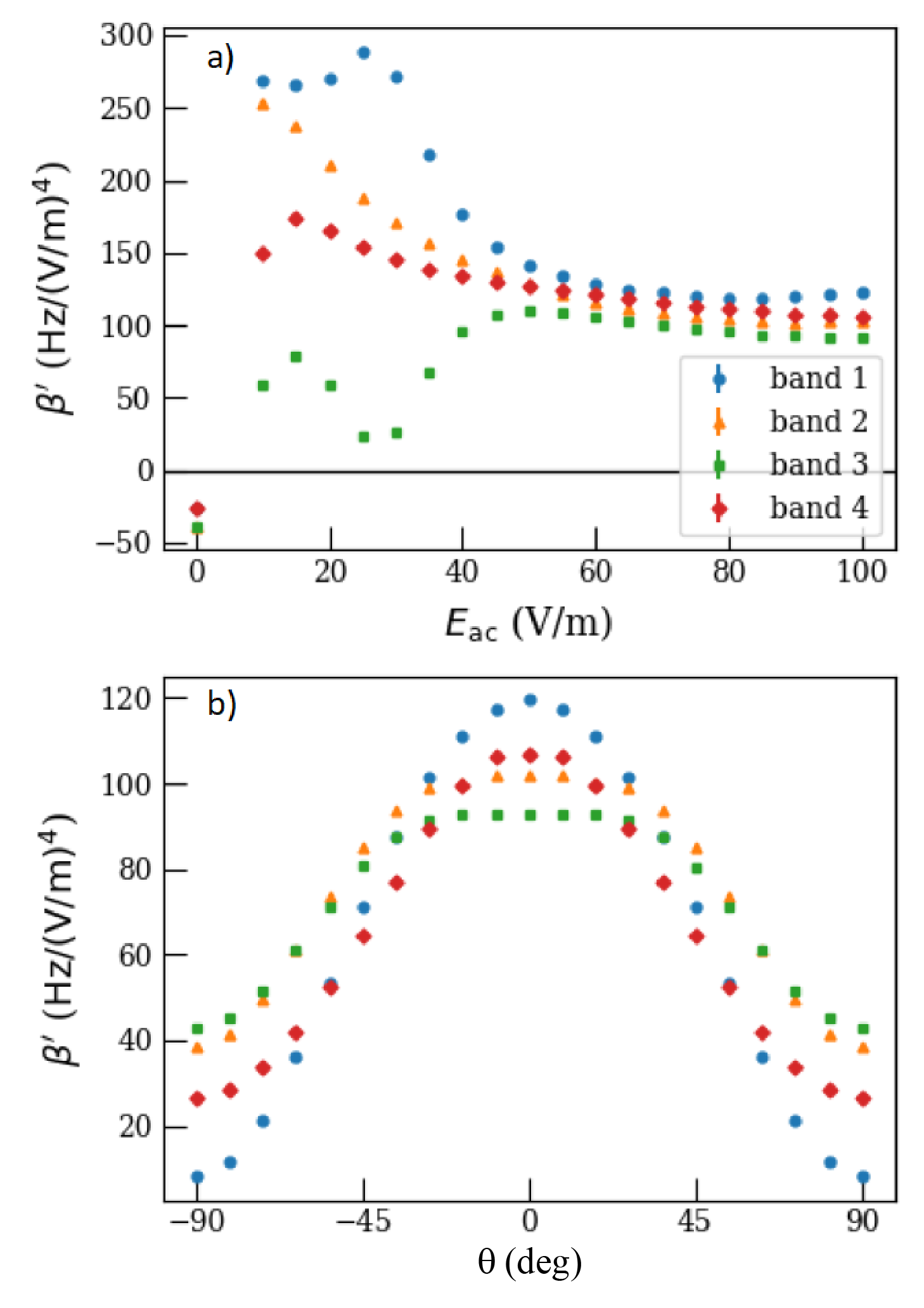}
    \caption{Hyper-polarizabilities for the dressed states in Fig. \ref{fig:Numerical Polarizability}. \textbf{a)} $\beta'$ dependence on dressing field strength. Sample at $5 ~\rm V/m$ is excluded due to a poor fit. \textbf{b)} Angular dependence of $\beta'$ at the band 4 nulling condition.}
    \label{fig:numerical beta}
\end{figure}

\begin{table}[!ht]
\begin{tabular}{|c|c|c|c|}
    \hline
     & Pin 1 & All-1 & All-2 \\
     \hline
    Band & $\beta'$ ($\rm MHz/V^4$) & $\beta'$ ($\rm MHz/V^4$) & $\beta'$ ($\rm MHz/V^4$)\\
    \hline
    Control &  N/A & N/A & N/A\\
    Dressed 1 & N/A & $0.240\pm0.12$ & $2.40\pm0.24$ \\
    Dressed 2 & N/A & $0.240\pm0.12$ & $2.40\pm0.24$ \\
    Dressed 3 & N/A & $0.216\pm0.144$ & $3.12\pm0.24$ \\
    Dressed 4 & N/A & $0.408\pm0.144$ & $1.92\pm0.24$ \\
    \hline
    \end{tabular}
    \caption{Hyper-polarizabilities in tested configurations. Control measurements on all pins and Pin 1 configuration data were fit to second-order polynomials so values are not available(N/A).}
    \label{tab:Beta}
\end{table}
Throughout this text numerical and experimental dc Stark-shift data has been fit to a fourth-order polynomial:

$$\Delta_{\rm dc}(E_{\rm dc}) = -\frac{1}{2}\alpha E_{\rm dc}^2 - \frac{1}{4!}\beta E_{\rm dc}^4$$

In the case of experimental data, the fit is instead parameterized by electrode voltages, that are linearly proportional to the electric field experienced by the atoms
$$\Delta_{\rm dc}(V_{\rm s}) = -\frac{1}{2}\alpha' (V_{\rm s}-V_0)^2 - \frac{1}{4!}\beta' (V_{\rm s}-V_0)^4.$$

The second-order coefficient $\alpha'$ is the familiar polarizability of the dressed states, while the fourth-order coefficient $\beta'$ is the hyper-polarizability of these states. This function was chosen for the dressed states, where the fourth-order shifts dominate over the sampled voltage ranges. The computed hyper-polarizabilities are presented here for both the numerical models and the experiment. In Fig. \ref{fig:numerical beta} we present the $\beta'$ coefficients that correspond to the $\alpha'$ data presented in Fig. \ref{fig:Numerical Polarizability}. In Fig. \ref{fig:numerical beta}a) we  see the fourth order effect is small in the absence of the dressing field, between -20 and -50 $\rm Hz/(V/m)^4$, but the sign and magnitude change rapidly with even small dressing fields present. At the band 4 nulling condition of $89.6 ~\rm V/m$, $\beta'$ varies between 80-150 $\rm Hz/(V/m)^4$, about 2-5 times the increase in the hyper-polarizabilities of the dressed states.

The experimental fourth-order fit coefficients are displayed in Table \ref{tab:Beta}. Major differences are notable between the two demonstrated configurations (All-1 and All-2). Sampling the Rydberg resonances at larger voltages would help to produce more reliable values for these coefficients.

The sudden rise in the hyper-polarizability when the dressing field is introduced indicates that for large dc fields, the electric field-sensitivity of the dressed states is significantly larger than the undressed case. There is a cross-over field strength, where the dressed dc Stark shift is larger than the un-dressed Stark shift. This seems to occur when the dc Stark matrix element between the dominant contributing states ( $ 52P_{3/2}$ and $ 51D_{5/2}$ in this case) is about the size of the Rabi frequency of the dressing field coupling between those two states. It stands to reason that the dressing field cannot reduce sensitivities to perturbations larger than the matrix elements it produces.

This also limits the practical uses of this scheme to cases where the dc field noise is small and the mean dc field is near zero. It also introduces a heuristic approach to finding the ideal detuning of the dressing field from the dressing transition, as a larger detuning will require a stronger dressing field to null the polarizability of the target state, which will in-turn, produce a smaller hyper-polarizability for the dressed state.

\section{Alternate Dressing Schemes} \label{Alternates}

\begin{figure}[!t]
    \centering
    \includegraphics[width=0.44\textwidth]{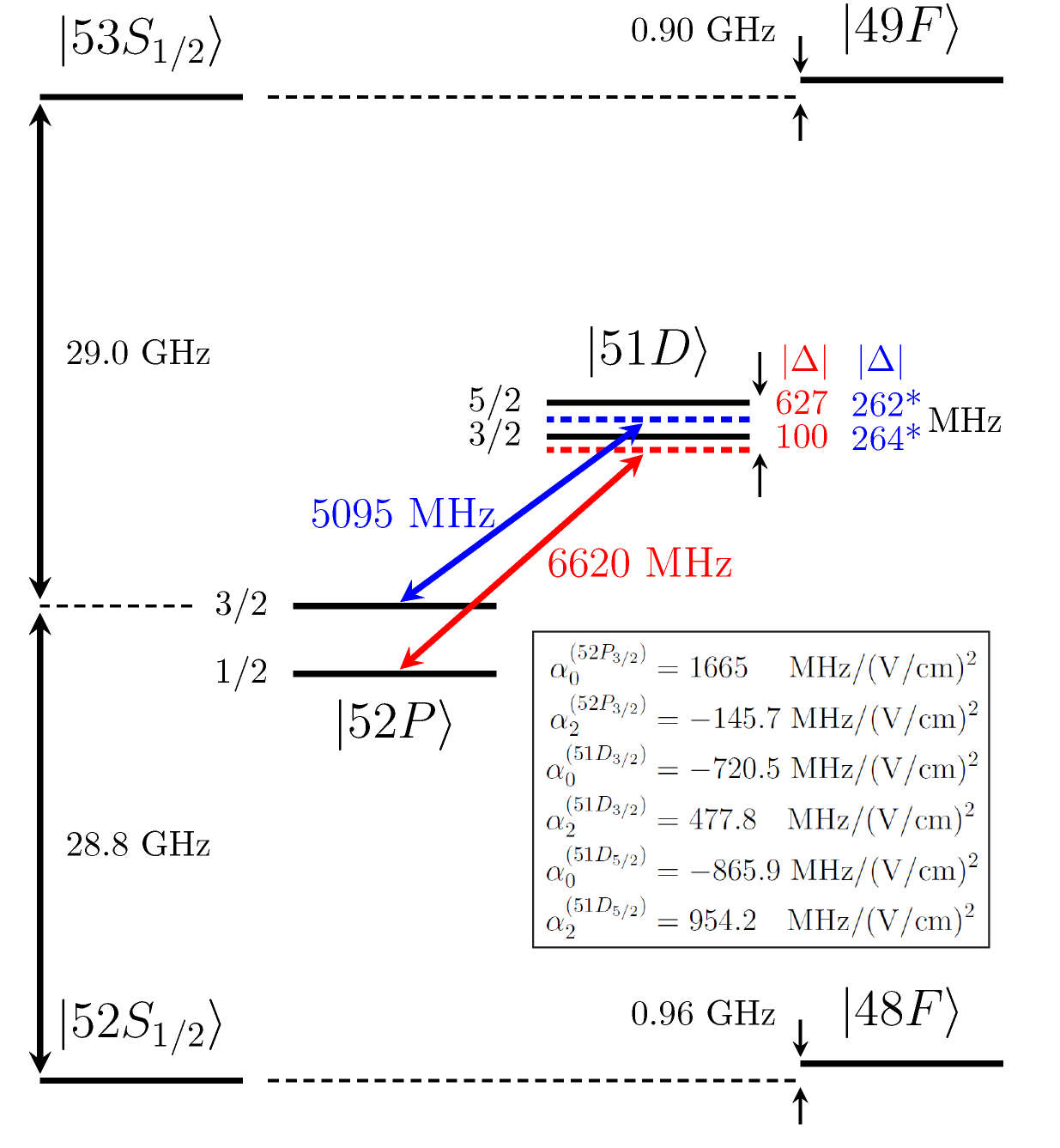}
    \caption{Energy level diagram for two 
    alternative dressing schemes. 1)  In the mid-detuning scheme the dressing field is tuned between the $ 52P_{3/2} - 51D_{3/2}$ transition and the $ 52P_{3/2} - 51D_{5/2}$ transition as  shown in blue. 2) In the $ 52P_{1/2}$ scheme, the dressing field is  red detuned from the transition $ 52P_{1/2} - 51D_{3/2}$, which is shown in red. The absolute value of field detuning from each transition is displayed on the right side of the $ 51D$ levels. \\
    * Detunings modified from ARC values, based on measurements presented in Appendix \ref{Resonance}}
    \label{fig:AltGrotrian}
\end{figure}

\begin{figure}[!t]
    \centering
    \includegraphics[width=0.45\textwidth]{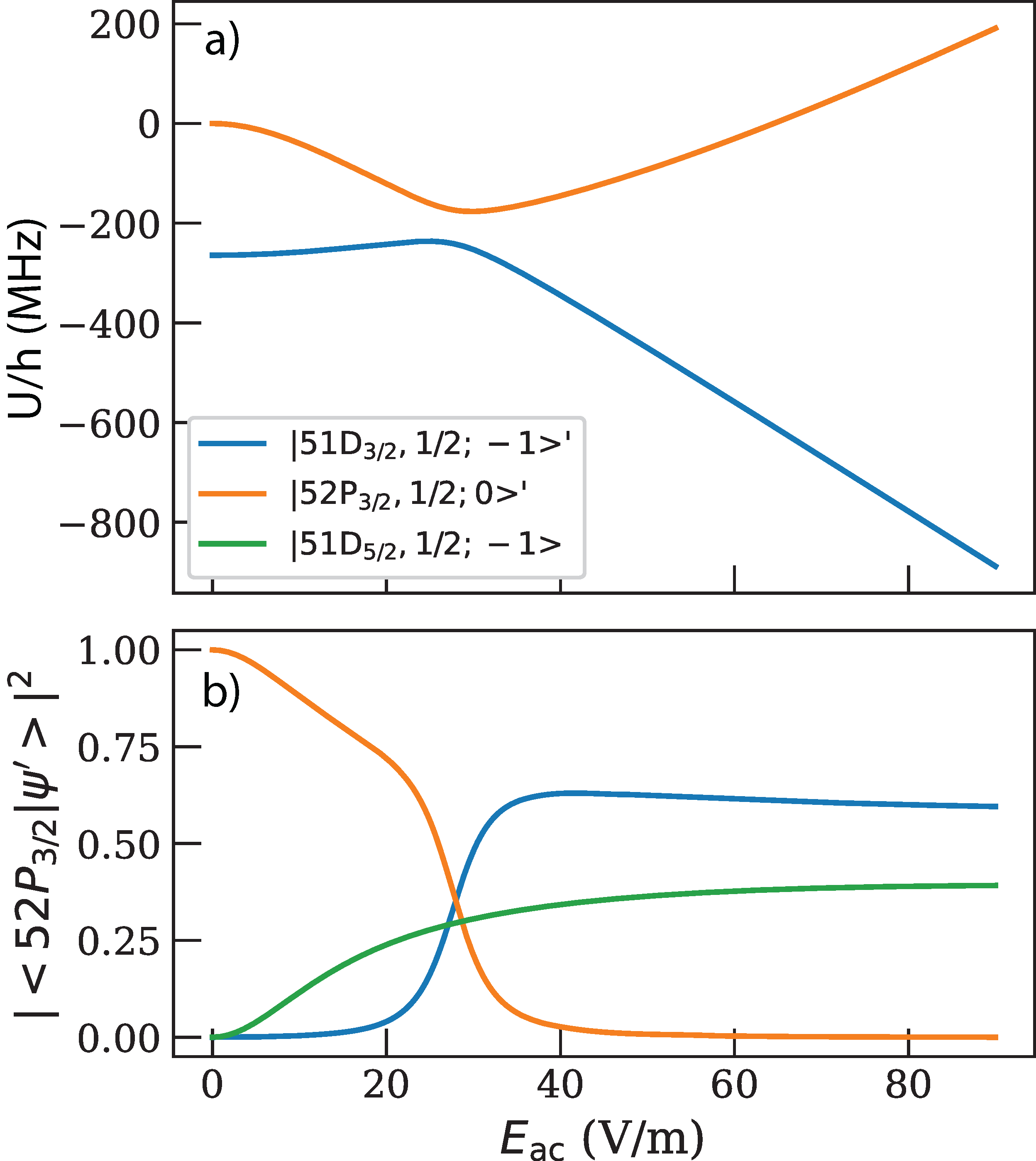}
    \caption{\textbf{a)} The energy spectrum of the $ 52P_{3/2}$ and $ 51D_{3/2}$, $m_j=1/2$ dressed states in the mid-detuning scheme. The avoided crossing is visible near $E_{\rm ac} = 25 ~\rm V/m$. \textbf{b)} Shows the square of the inner-product between dressed states and the bare $ \ket{52 P_{3/2},1/2;0}$ state. The high levels of mixing between the $P$ and $D$ states are evident near the avoided-crossing.}
    \label{fig:AvoidedCrossing}
\end{figure}

In addition to the red-detuned dressing scheme presented in the main text, two alternative schemes for polarizability reduction have been considered numerically. These schemes are presented with the ellipticity of the dressing field set to zero, for ease of demonstration and analysis. In this case only half of the magnetic angular momentum states in the Hilbert space are plotted ($m_j > 0$), as there are no matrix elements between different $m_j$ states, and the behavior of the positive and negative $m_j$ states is identical.

The first of the two additional schemes presented here namely called ``Mid-Detuning" scheme. In this scheme the dressing tone of $\omega_d = 2\pi\times5095~\rm MHz$ admixes the $ 52P_{3/2}$ states with the $ 51D_{3/2}$ and $ 51D_{5/2}$ states. This tone lies between the $ 52P_{3/2}-51D_{3/2}$ and $ 52P_{3/2}-51D_{5/2}$ resonances. This scheme proved interesting because there is an avoided crossing between the $\ket{ 52P_{3/2},\pm 1/2;0}$ and $\ket{51D_{3/2},\pm1/2;-1}$ states as the dressing field amplitude increases.  At this avoided crossing there are high-levels of mixing between the $P$ and $D$ states, resulting in a few nulling conditions for the dressed states, including one at a lower field amplitude, which may be advantageous in some systems.

The second scheme explored was the ``$ 52P_{1/2}$'' scheme, where the $ 52P_{1/2}$ states are admixed with the $ 51D_{3/2}$ states coupled with a red-detuned dressing field at $\omega_d=2\pi\times6620 ~\rm MHz$. This scheme was explored with the expectation that the dressed Rydberg states will have a more isotropic polarizability than their $\ket{ 52P_{3/2}}$ counterparts, due to the simpler angular momentum structure of the $ 52P_{1/2}$ level.

\subsection{Mid-detuning}

\begin{figure*}[!t]
    \centering
    \includegraphics[width=\textwidth]{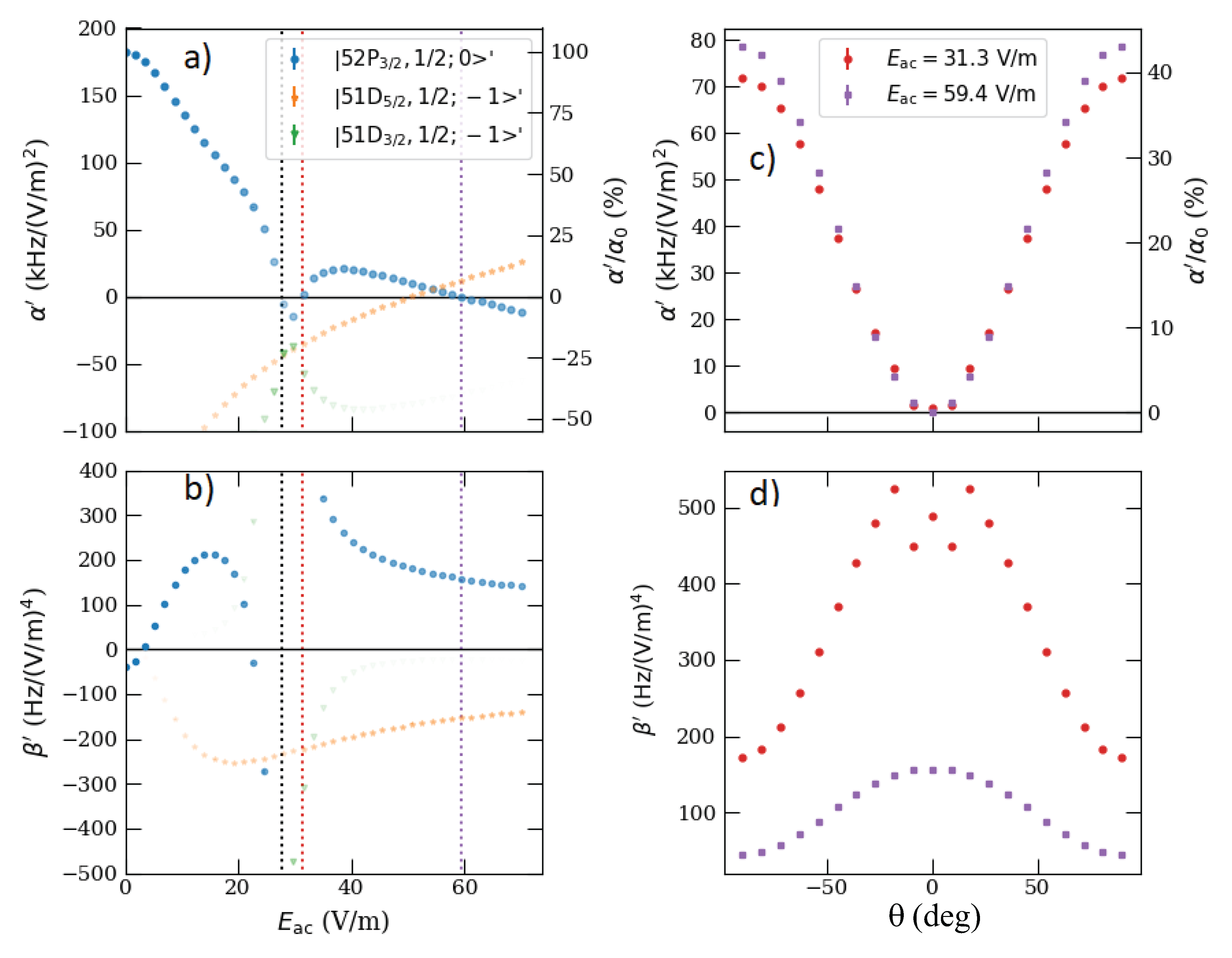}
    \caption{Polarizability and Hyper-polarizability values for the dressed states and their anisotropy  in the mid-detuning scheme. \textbf{a)} Polarizabilities of relevant Rydberg states, and their square-overlap with the un-dressed $\ket{ 52P_{3/2},1/2}$ state represented by their opacity. The color in this case indicates which un-dressed Rydberg state has the greatest overlap with the corresponding dressed-state. Three zero-crossings are visible for the $ 52P_{3/2}$ dressed state, indicated by vertical lines in black, red and purple from left to right corresponding to $E_{\rm ac}= 27.2 ~\rm V/m$, $30.55~\rm V/m$, and $60.4~\rm V/m$  respectively. \textbf{b)} Corresponding hyper-polarizability values for the states shown  in a). There is a large increase in hyper-polarizabilty near the avoided crossing. \textbf{c)} Polarizability values for the second and third (red and purple, dotted lines) nulling condition. Both nulling conditions show small improvements in anisotropy compared to the dressing scheme explored in the main text, with the dressed states retaining only $\sim 40\%$ of the undressed polarizability.  The first nulling condition is not explored for anisotropy due to the divergent hyper-polarizability. \textbf{d)} Corresponding hyper-polarizability values for the nulling conditions in  c). The 31.3 V/m nulling condition has five times larger hyper-polarizability compared to the scheme presented in the main text, while the 59.4 V/m nulling condition has comparable hyper-polarizability to that of the main text.}
    \label{fig:MidPol}
\end{figure*}

In this scheme the dressing tone at $\omega_d = 2\pi\times5095~\rm MHz$ is blue-detuned from the $ 52P_{3/2}-  51D_{3/2}$ transition and red-detuned from the $ 52P_{3/2}- 51D_{5/2}$ transition. Of interest in this discussion, as the dressing field amplitude increases, an avoided crossings occurs between the $\ket{ 52P_{3/2},1/2;0}$ and $\ket{51D_{3/2},1/2;-1}$ dressed states. These avoided crossings are presented in Fig. \ref{fig:AvoidedCrossing}. We  explored using the high levels of mixing that occur at this avoided crossing, to find a polarizability nulling condition nearby.

 The avoided crossing occurs due to the influence of the $ 52P_{1/2}$ states on the $ 51D_{3/2}$ states, as the tone is red-detuned from that transition, causing the $ 51D_{3/2}$ states to be ac-Stark shifted to higher energies. The $ 52P_{1/2}-51D_{3/2}$ matrix element is  much larger than the $ 52P_{3/2}-51D_{3/2}$  matrix element, resulting in a net upward energy shift of the state. A similar pattern is followed by the $ 52P_{3/2}$ state, being ac-Stark shifted down by the coupling to  $ 51D_{5/2}$. As the dressing field strength increases,  the two dressed states are shifted towards the same energy, but are deflected due to their non-zero matrix element. 

In order to investigate the effects of this avoided crossing on the dressed state polarizabilities, the procedure described in Sec. \ref{Numerical} was followed for dressing field amplitudes between $E_{\rm ac} = 0 ~\rm V/m$ and $E_{\rm ac} = 70 ~\rm V/m$. The resulting polarizability values were then interpolated with a quadratic spline and the zero-crossings of the function were found. The polarizability and hyper-polarizability of the dressed states in this scheme are presented in Fig. \ref{fig:MidPol} a,b.

Three nulling conditions were found $E_{\rm ac}= 27.2 ~\rm V/m$, $30.55~\rm V/m$, and $60.4~\rm V/m$. The first two nulling conditions are found near the avoided crossing while the last one is due to the mixing of the $ 52P_{3/2}$ and $51D_{5/2}$ levels. The first nulling condition occurs at a point where the hyper-polarizability of the dressed states is seemingly diverging, so it is not explored further. The other two nulling conditions are investigated for anisotropy was studied and shown in Fig. \ref{fig:MidPol} c,d.

Both nulling conditions show large amounts of anisotropy, retaining around 40\% of their undressed polarizability, which is a small  improvement over the scheme presented in the main text. The $E_{\rm ac} = 30.55 ~\rm V/m$ nulling condition shows a very large increase in the hyper-polarizability compared to the $E_{\rm ac}=60.4 ~\rm V/m$ condition, and the scheme presented in the main text. With a hyper-polarizability of over $500 ~\rm Hz/(V/m)^4$, compared to the $120 ~\rm Hz/(V/m)^4$ and $160 ~\rm Hz/(V/m)^4$ values from the main text and $E_{\rm ac}=60.4 ~\rm V/m$ condition respectively. This limits the utility of the avoided-crossing based polarizability nulling even further to systems with smaller dc electric field noise.

\subsection{$ 52P_{1/2}$ scheme}
\begin{figure}[t]
    \centering
    \includegraphics[width=0.5\textwidth]{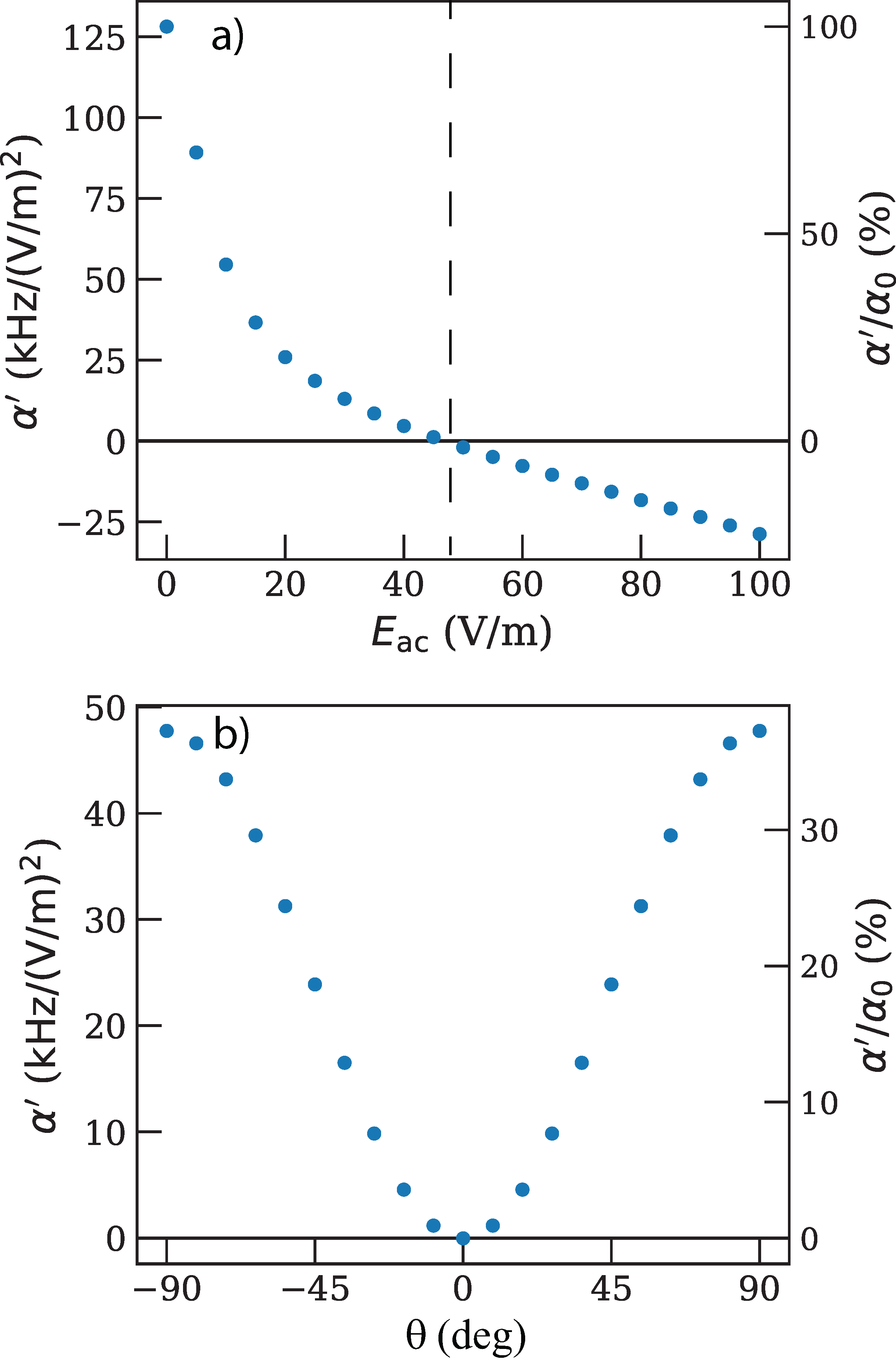}
    \caption{Polarizabilities of the $ 52P_{1/2}$ states when dressed with a tone at $\omega_d=2\pi\times6620~\rm MHz$. \textbf{a)} Polarizability as a function of the dressing field amplitude, the second order nulling condition is found when $E_{\rm ac}=46.8~\rm V/m$ (indicated by the vertical dashed line). \textbf{b)} Anisotropy of $\alpha'$ at the nulling condition. The reduced polarizability in the worst case is less than 40\% of the un-dressed value, an improvement over the $ 52P_{3/2}$ schemes interrogated.}
    \label{fig:low alpha}
\end{figure}

\begin{figure}[t]
    \centering
    \includegraphics[width = 0.5\textwidth]{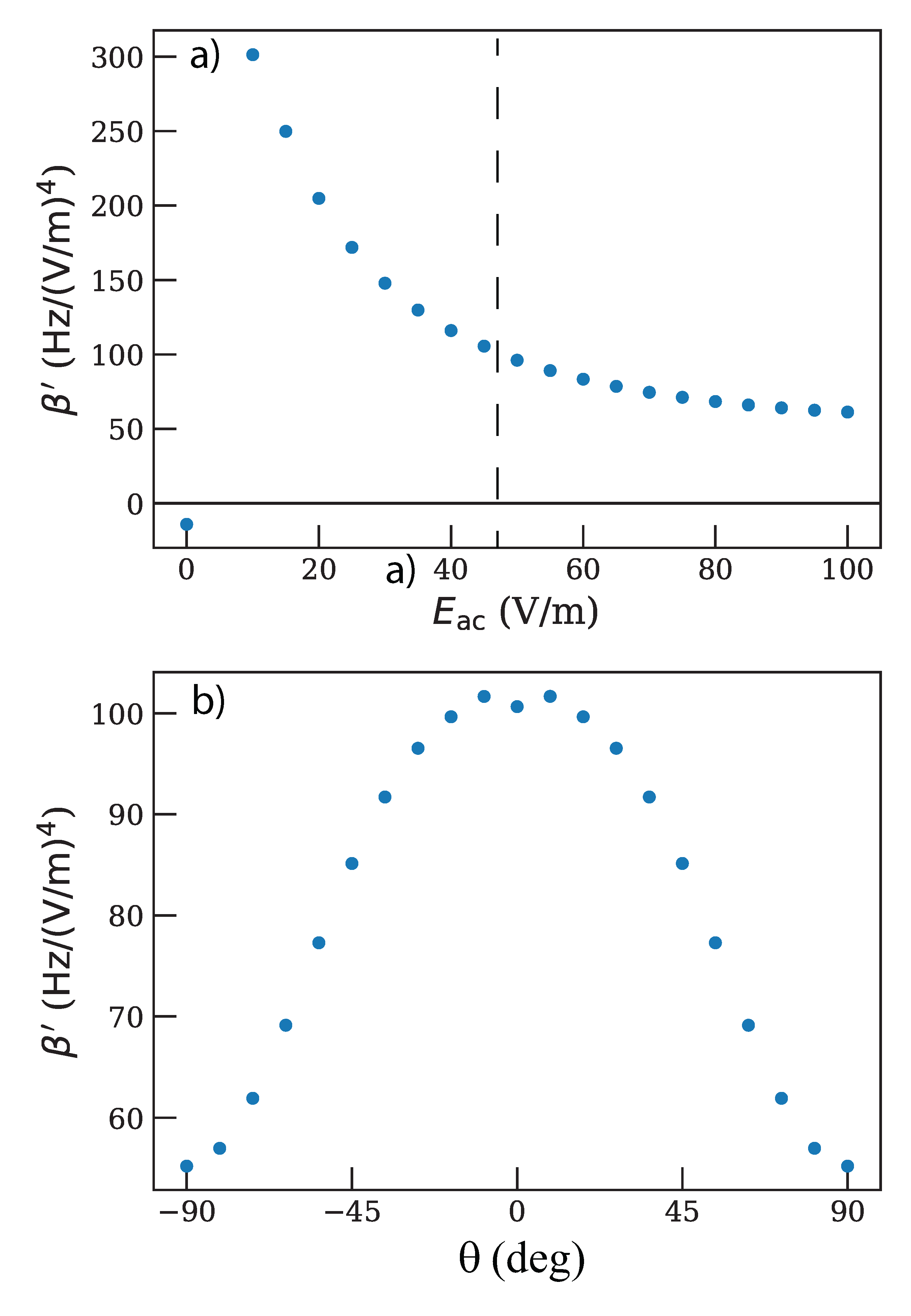}
    \caption{The corresponding hyper-polarizability values to those plotted in Fig.\ref{fig:low alpha}. \textbf{a)} The hyper-polarizability is seen to increase at small values of $E_{\rm ac}$, reducing in magnitude slowly as the dressing field power increases. The value at the second-order nulling condition (indicated by the vertical dashed line) is similar in magnitude to the $ 52P_{3/2}$ case in the main text. \textbf{b)} The anisotropy of the hyper-polarizability at the nulling condition}
    \label{fig:low beta}
\end{figure}

In this scheme the goal is to reduce the polarizability of the $52P_{1/2}$ states by admixing the $51D_{3/2}$ states. The dressing field frequency is red-detuned from the $ 51P_{1/2}-51D_{3/2}$ transition by $100 ~\rm MHz$. The expectation was that there would be a reduced anisotropy from the simpler angular momentum structure of $J=1/2$.

The numerical results are presented in Figs. \ref{fig:low alpha} and \ref{fig:low beta}. The dressing field was modeled with parameters $\omega_d = 2\pi\times6620$ MHz and $\epsilon=0$, and the nulling condition was found at $E_{\rm ac}=46.8 ~\rm V/m$. The anisotropy is significantly lower than in any of the $52P_{3/2}$ schemes, with 35\% of the original polarizability remaining when the fields are orthogonal. The hyper-polarizability of the nulled states is comparable to that of the scheme considered in the main text.

\section{Ground State Autler-Townes Doublets} \label{AT}
\begin{figure}[!t]          
    \centering
    \includegraphics[width=0.49\textwidth]{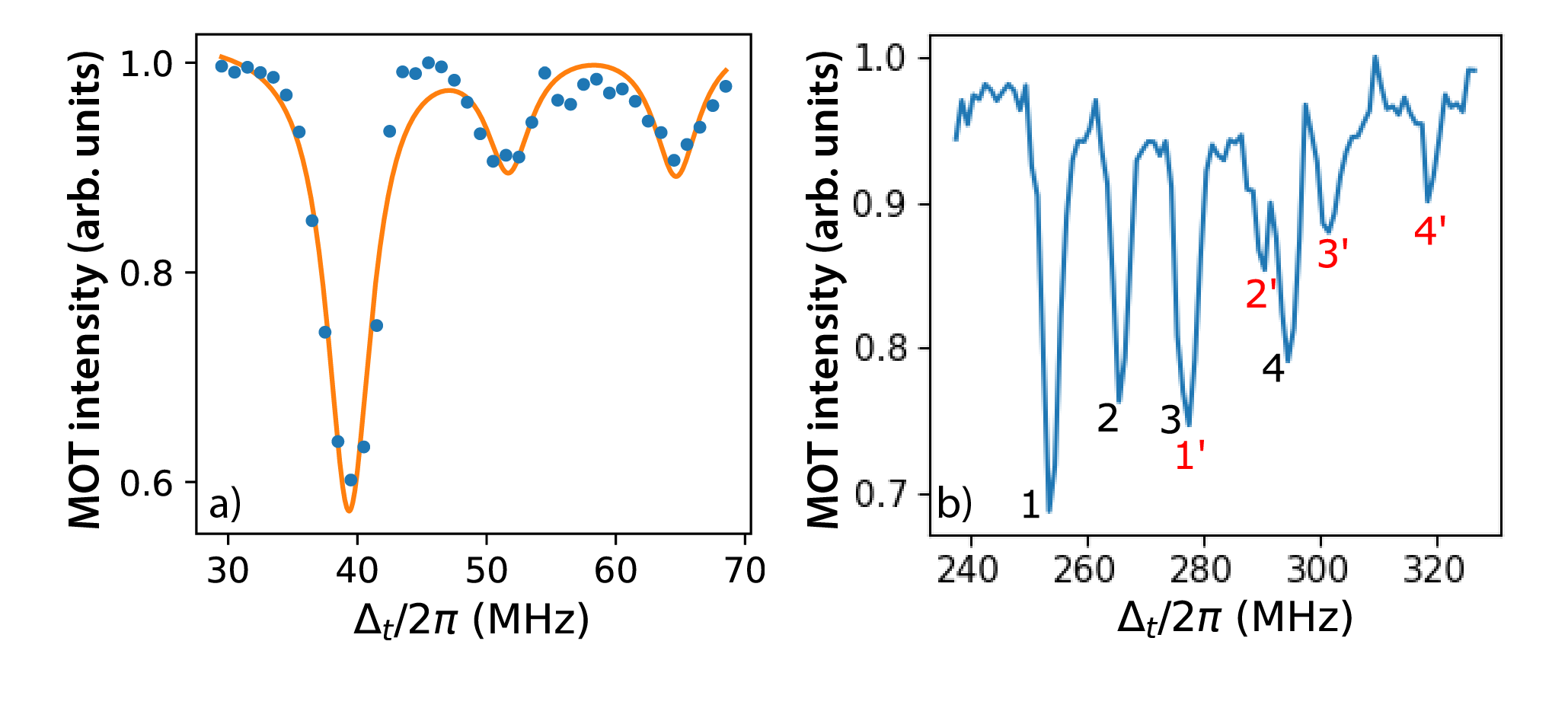} 
    \caption{\textbf{a)}  Sample MOT depletion data where the Autler-Townes spectra of the ground state are visible. Larger offset frequencies correspond to lower frequencies of the 595 nm laser. Line centers are at $\mu_0=39.4 \pm 0.10 ~\rm MHz$, $\mu_1=51.7 \pm 0.38 ~\rm MHz$ and $\mu_2=64.7 \pm 0.34 ~\rm MHz$. The deepest line at 39 MHz corresponds to the ac Stark-shifted two-photon resonance while the two additional features are due to the Autler-Townes splitting of the ground state from the 852 nm MOT trapping laser and the 685 nm Rydberg laser respectively. These data were taken with the 685 nm laser  80 MHz blue detuned from the $\ket{ 6S_{1/2},F=4}- \ket{5D_{5/2},F'=6}$ quadrupole transition. \textbf{b)} An example of MOT depletion spectra where AT bands are visible. Taken with the pump beam 80 MHz blue detuned of the $\ket{6S_{1/2},F=4} - \ket{5D_{5/2},F'=6}$ transition, $\omega_d = 2\pi\times4780 ~\rm MHz$, and  $P_{\rm s}=-8.5 ~\rm dBm$. AT doublets are labeled numerically with 
     the actual resonances  denoted in black (labeled as 1-4), and their corresponding AT resonances denoted in red (labeled as 1'-4'). }
    \label{fig:AT}
\end{figure}
The ladder excitation scheme depicted in Fig. \ref{fig:Excitation} can be described as a pump (685 nm) and probe (595 nm) type spectroscopy scheme. The Rydberg resonance signals are seen when the two-photon resonance condition is met
$$\omega_1 + \omega_2 = \omega_{\rm Rg}+\Delta_{\rm ac}$$
Where $\Delta_{\rm ac}$ is the differential ac Stark-shift induced on the ground to Rydberg state transition due to the pump and probe beams\cite{Maller2015}. The Rabi coupling, due to the 685 nm pump field, between the ground and $ 5D_{5/2}$ states creates Aulter-Townes (AT) doublets of the ground state and of the 5D states. The energy splitting between the two resonances is \cite{Teo2003} 
$$|\Delta_{\rm AT}| = \sqrt{\Omega_1^2+\Delta^2} \equiv \Omega_1'$$
and the ratio of the amplitudes of the resonances in the doublet is
$r_{\rm AT} = \tan^4(\theta)$
where
$ \cos^2(\theta)= \frac{1}{2}\left(1+\Delta/\Omega\right).$
The energy splitting of the AT doublet on the ground state can be probed by setting the probe frequency to match the two-photon resonance condition of the additional AT resonance
$$\omega_1 + \omega_2 = \omega_{\rm Rg}+\Delta_{\rm ac} \pm |\Delta_{\rm AT}|$$
where the sign is determined by the sign of the pump detuning $\Delta$. In the case of a blue (red)-detuned pump, the ground state will see a positive (negative) ac Stark-shift and the AT resonance will occur at a greater (lesser) probe frequency, thus the positive (negative) sign will be taken.

In this excitation scheme the pump laser is blue-detuned from the $\ket{ 6S_{1/2},F=4}-\ket{5D_{5/2},F'=6}$  transition, while red-detuned from the $\ket{6S_{1/2},F=4}-\ket{5D_{5/2},F'=5}$  transition. In this case two AT resonances become visible on either side of the actual resonance. The splitting of each AT band corresponding to the actual resonance remains the generalized Rabi frequency of the particular transition. 

Experimentally the formation of the AT splitting became problematic at larger dressing powers, as the splitting of the dressed Zeeman bands into AT doublets caused difficulty in  correctly mapping the dressed Zeeman bands. This was due to the comparable MOT depletion amplitude of the  dressed Zeeman bands and their AT resonances. This is evident in Fig. \ref{fig:AT}b, where eight resonant features are visible at high dressing field powers, with the actual resonances  denoted in black (labeled as 1-4), and their corresponding AT resonances denoted in red (labeled as 1'-4').

Initially, the 685 nm laser was $80 ~\rm MHz$ blue detuned to the $4-6'$ quadrupole transition. This resulted in the spectrum shown in Fig. \ref{fig:AT}, the main resonance can be seen at 39.4 MHz, and the AT resonance due to the $\rm 4-5'$ detuning is visible at 64.7 MHz. The energy splitting of the doublet is $2(64.7-39.4) ~\rm MHz$ $=50.6~\rm MHz$, while the $\rm 4-5'$ detuning is 47.2 MHz. Based on the splitting the $\rm 4-5'$ Rabi frequency is estimated to be $\Omega_{45'} \sim 18.3$ MHz.

The AT resonance issue was ameliorated by changing the pump frequency to be 60 MHz blue detuned of the 4-6' transition (as in Fig. \ref{fig:Excitation}). Increasing the ratio $\Delta_{45'}/\Omega_{45'}$  increased the separation of  the AT band from  the actual resonances of interest. Further increasing $\Delta_{45'}$ reduced the separation  of the $\rm 4-6'$ AT band to the point where it interfered with the spectroscopy as the $\rm 4-5'$ resonance observed previously.

The MOT trapping beams also formed an AT doublet \cite{JBai2020}, as shown in Fig. \ref{fig:AT} at $51.7 ~\rm MHz$. This band proved to not be an issue at high dressing powers. The AT splitting of this feature is $2(51.7-39.4)~\rm MHz$=$24.6~\rm MHz$. The MOT beams are $8.9$ MHz red detuned  to the $\ket{ 6S_{1/2},F=4}- \ket{ 6P_{3/2},F'=5}$ transition, thus we expect the MOT beam Rabi frequency $\Omega_{\rm MOT}=2\pi\times 11.6$ MHz.
\section{Measurement of $52P_{3/2}-51D_{3/2}$ resonance} \label{Resonance}
\begin{figure}[!t]
    \centering
    \includegraphics[width=0.48\textwidth]{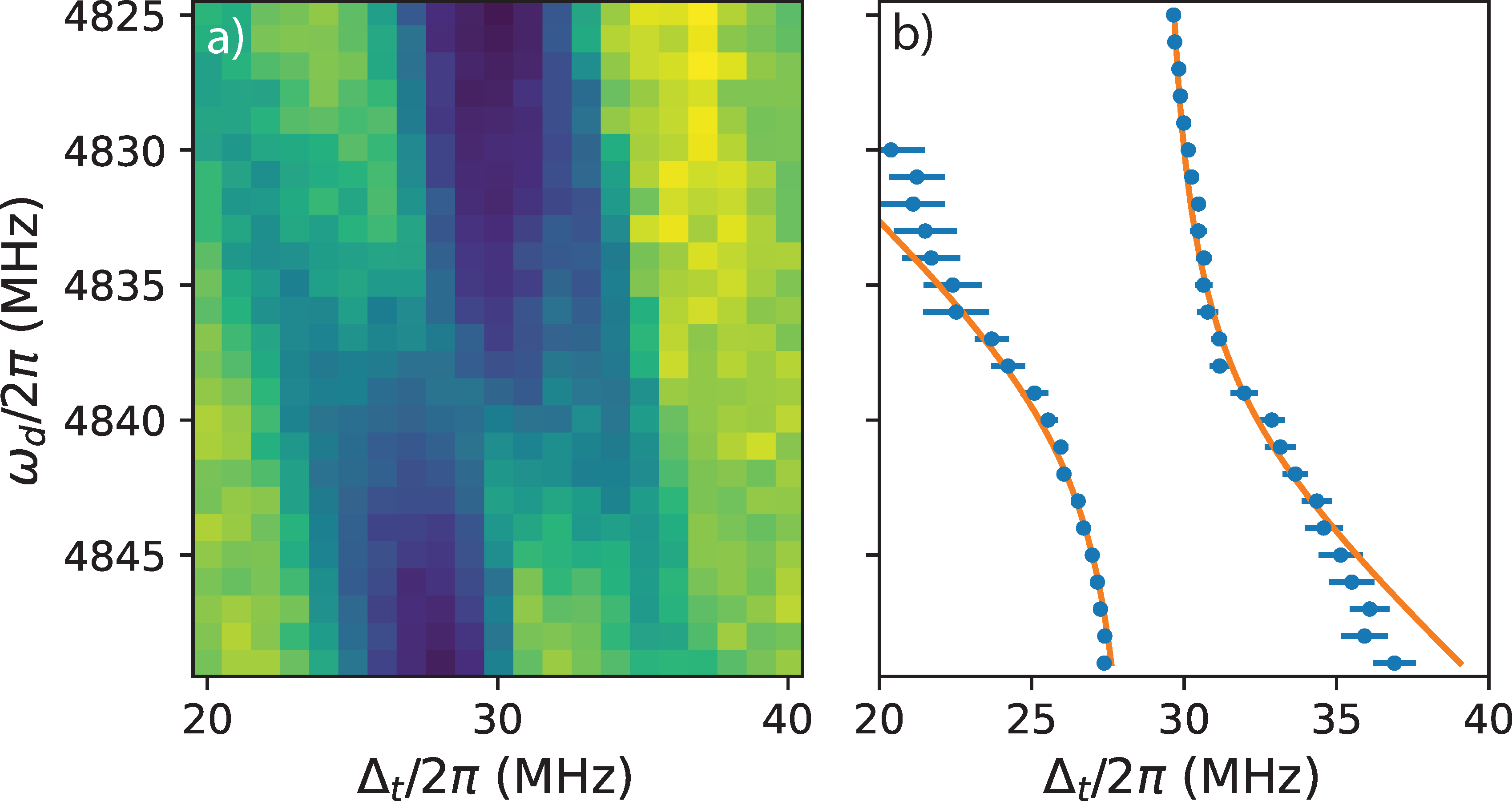}
    \caption{Experimental Measurement of the $ 52P_{3/2} - 51D_{3/2}$ resonance. A low power dressing tune was used ($P_s=-40~\rm dBm$) and the dressing frequency was swept across the observed Rydberg-Rydberg resonance. The resonance can be seen above as the avoided crossing of the two lines. \textbf{a)} The raw MOT depletion spectroscopy data. \textbf{b)} The resonances are fit to the characteristic curves of an avoided crossing  and the Rydberg-Rydberg resonance is computed from the fit. A resonance of $\nu_{\rm res} = 4840.1 \pm 0.3 ~\rm MHz$ was computed, a 10 MHz difference from the value predicted by ARC.}
    \label{fig:Ryd-Ryd}
\end{figure}

We measured the $ 52P_{3/2}-51D_{3/2}$ transition frequency using AT splitting spectroscopy. For this measurement we set $P_{\rm s}= -40~\rm  dBm$ and the dressing field frequency  was swept across the resonant frequency, as computed using ARC. This spectroscopy resulted in  avoided crossing resonances as shown in Fig. \ref{fig:Ryd-Ryd}. The measured energy spectrum is in good agreement with the calculated AT splitting $E_{\pm}(\nu)$ using 
\begin{align*}
    E_{\pm}(\nu) &= \frac{\nu-\nu_{\rm res}}{2}\pm\frac{\sqrt{\left(\nu-\nu_{\rm res}\right)^2+\Omega^2}}{2}+\nu_{\rm bare}
\end{align*}

The resulting fit predicts a Rydberg-Rydberg resonance of  $\nu_{\rm res} = 4840.1 \pm 0.3 ~\rm MHz$, about $10 ~\rm MHz$ different from the ARC value. This difference may be caused by ac-Stark shifts on the Rydberg levels from the Rydberg and trapping laser fields. 

There is a disagreement between the fit curves and the ground-Rydberg resonances at large detunings, which may be due to a frequency-dependent amplitude of the dressing field. The coaxial cable carrying the dressing signal into the chamber is not impedance matched on the feed-through or antenna ends, resulting in standing waves in the line. When the wavelength of the dressing field matches harmonics of the coaxial line, the standing wave results in a larger Rabi frequency at the atoms. We have measured the resonance structure using a vector network analyzer, and one of these resonances is observed near $\sim$ 4840 MHz. The discrepancies observed at the larger detunings do not, however,  appreciably affect the reported fit results.

\section{Center of Gravity (CoG) subtraction} \label{CoG}

\begin{figure}[!t]
    \centering
    \includegraphics[width=0.45\textwidth]{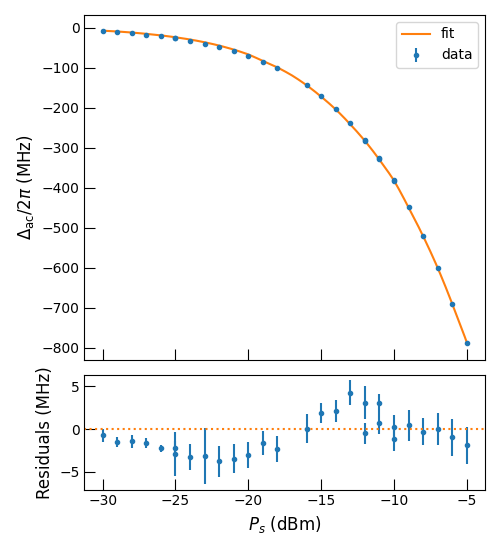}
    \caption{Measured ac Stark shift data of the $52P_{3/2}$ level averaged over the $m_j$ states  as a function of dressing field power. The  bottom figure shows the residuals deviated from the calculated values.}
    \label{fig:CoG Fit}
\end{figure}

In Fig. \ref{fig:RF calibration} experimental and numerical spectroscopy data is presented, demonstrating the ac-Stark shifts of the dressed Rydberg energy levels. These data are presented with respect to estimates of the mean resonance frequency of all four Zeeman bands at a given dressing power. We refer to the  mean resonance frequency as the Center of Gravity (CoG) frequency for the $ 52P_{3/2}$ level
\begin{align} \label{eq.CoG}
    \omega_{\rm CoG}(P_s) = \frac{1}{4}\sum_{i=1}^4 \omega_i(P_s)   
\end{align}
with $\omega_i(P_s)$  the band specific resonance frequencies.

This quantity allowed scans of the Rydberg laser frequencies to be conveniently defined as the dressing power was varied. The necessity for this can be seen in Fig.\ref{fig:RF calibration} e), where the ac-Stark shifts on all the bands spanned about $800 ~\rm MHz$, while the inter-band spacing only spanned about $160 ~\rm MHz$. In order to   conveniently analyze  scans of the Rydberg frequency and the dressing field power, we estimated the CoG Rydberg resonance as a function of power as 
\begin{align} \label{eq.xi}
    \mu_e(P_s) &= \mu_{\rm bare}-\xi_{\rm ac}\left(\sqrt{c_0^2+10^{P_{\rm s}/10}}-c_0\right).
\end{align}
This expected detuning is parameterized by $\mu_{\rm bare}$, the undressed ground-Rydberg resonance frequency, $\xi_{\rm ac}$ an effective ``ac-polarizability" parameter, with units of frequency, and the unit-less $c_0$ parameter. For the experimental data presented in Fig. \ref{fig:RF calibration}a)-d), the values of $\xi_{\rm ac}$, $c_0$  
were $-854.17 ~\rm MHz$ and $0.1138$  respectively,  which were obtained from fitting to initial ac Stark shift measurements similar to those shown in Fig. \ref{fig:CoG Fit}.

 The detuning of the 1190 nm laser, from this expected CoG resonance ($\Delta_{\rm e} = \Delta_{\rm t}-\mu_e(P_s)$) was scanned in conjunction with the power. This allowed us to detect the fine motion of the four bands, while also probing the larger scale Stark shifts induced by the dressing field. The value of $\Delta_{\rm t}$ was stored for each data-point in the scans, and was used when computing the ac-Stark shifts of each band, these ac-Stark values were used to compute the fit.
 
 The data in the inset of Fig. \ref{fig:RF calibration}e) is presented with an estimated CoG resonance subtracted out of both the numerical and experimental data. The same function is subtracted out of both datasets. This $\mu_e(P_s)$ was newly computed using the experimental ac-Stark shift data presented in Fig. \ref{fig:RF calibration}a)-d), retaining the same functional form as above. Values of $\xi_{\rm ac} = -654 ~\rm MHz$ and $c_0=0.292$ were used in this computation. These values were derived using a least-squares fit to the experimental ac-Stark shift data presented in Fig. \ref{fig:CoG Fit}. 

\end{document}